\documentclass[aps,prb,showpacs,superscriptaddress, preprint numbers,twocolumn]{revtex4-1}
\usepackage{rotating}
\usepackage{epstopdf}
\usepackage{color}
\usepackage{amsmath}
\usepackage{graphicx}
\usepackage{hyperref}
\usepackage{color}
\usepackage{amssymb}
\usepackage{dcolumn}
\usepackage{bm}
\usepackage{epsfig}
\usepackage{array}
\usepackage{subfigure}
\usepackage{upgreek}
\usepackage{wasysym}
\usepackage{slashed}
\usepackage[section]{placeins}
\usepackage{appendix}
\usepackage{subfigure}
\usepackage{multirow}
\usepackage{adjustbox}
\newcommand{\be}{\begin{equation}} 
\newcommand{\ee}{\end{equation}} 
\newcommand{\bea}{\begin{eqnarray}} 
\newcommand{\eea}{\end{eqnarray}} 
\newcommand{\bqa}{\begin{eqnarray}}
\newcommand{\eqa}{\end{eqnarray}}

\newcommand{\w}{\omega}

\newcommand{\figphononA}{
\begin{figure}[htb]
\centering
\hspace{0cm}
\subfigure[noonleline][]
{\label{fig: ac_phonon_memory_full}\includegraphics[height=35mm,width=40mm]{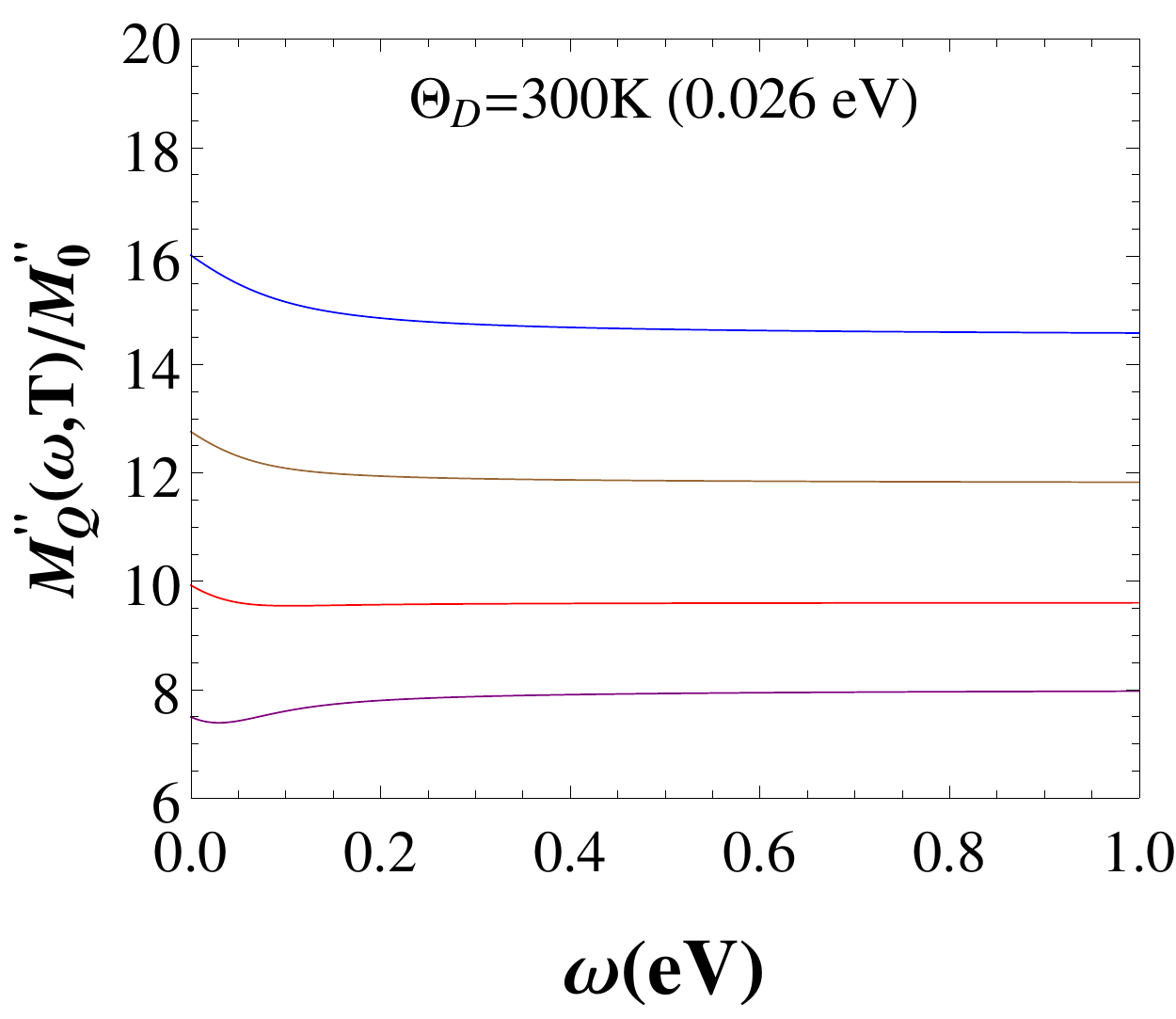}}
\hspace{0cm}
\subfigure[noonleline][]
{\label{fig: ac_phonon_memory_half}\includegraphics[height=35mm,width=40mm]{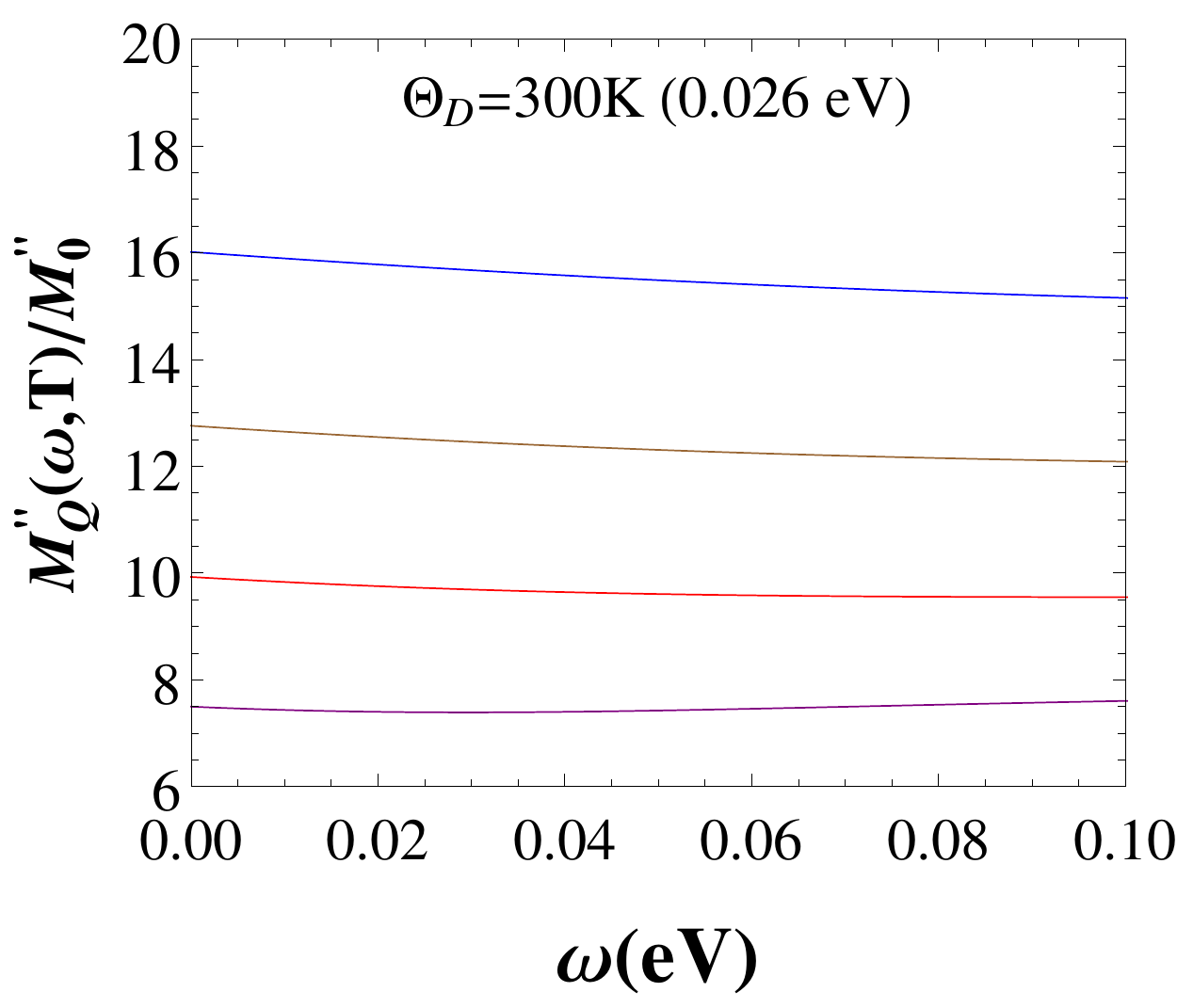}}
\caption{(a): Plot of imaginary thermoelectric memory function for phonon case at different temperatures such as $200$ K (0.017 eV, purple), $250$ K (0.021 eV, red), $300$ K (0.026 eV, brown), $350$ K (0.030 eV, blue). (b): The low frequency regime of Fig.(\ref{fig: ac_phonon_memory_full}).}
\label{fig: ac_phonon_freq}
\end{figure}
}
\newcommand{\figphononB}{
\begin{figure}[htb]
\centering
\includegraphics[height=40mm,width=60mm]{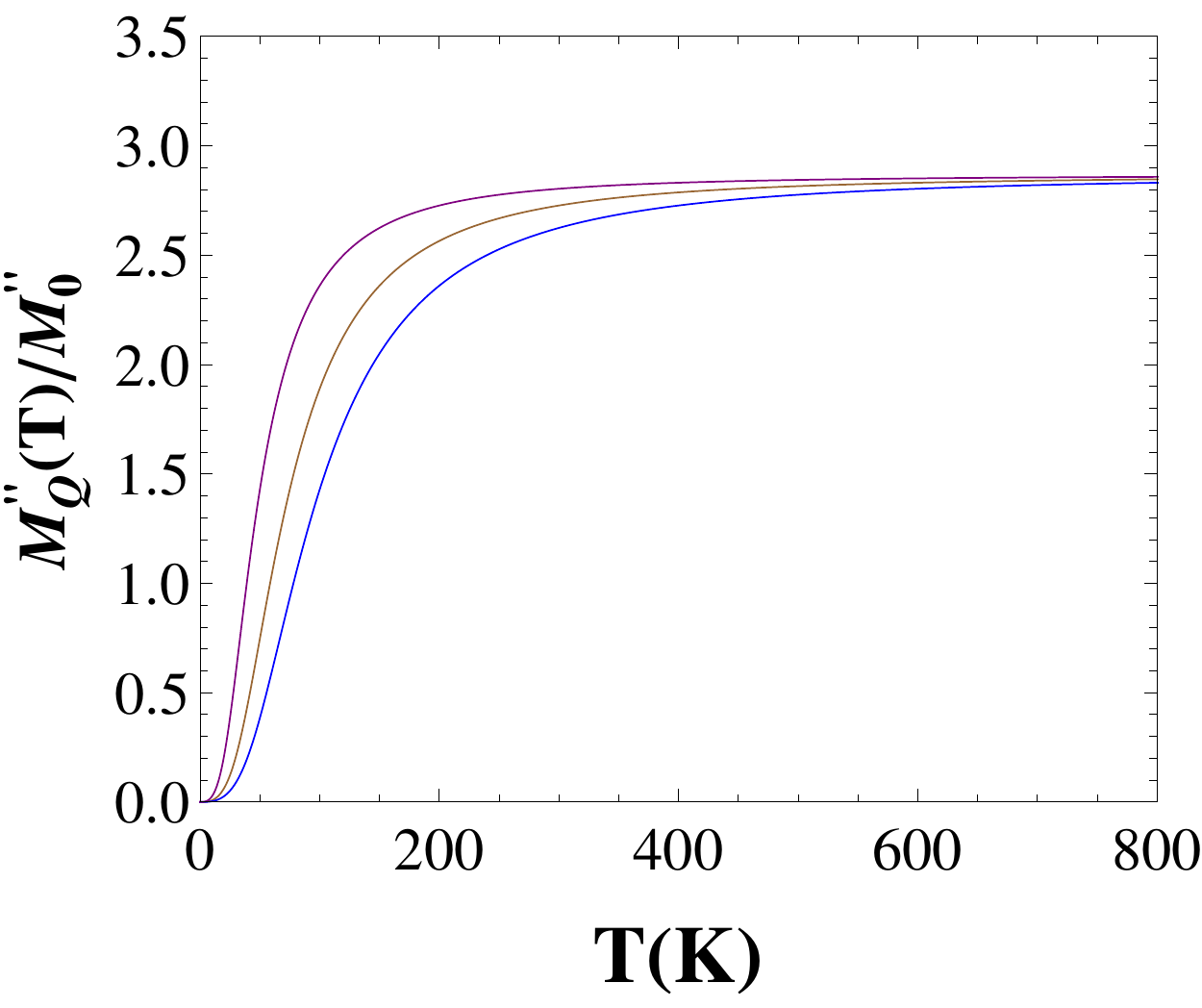}
\caption{Plot of imaginary part of the dc thermoelectric memory function for phonon case at different Debye temperatures such as $200$(purple), $300$(brown) and $400$K(blue).}
\label{fig: dc_phonon}
\end{figure}
}
\newcommand{\figphononC}{
\begin{figure}[htb]
\centering
\hspace{0cm}
\subfigure[noonleline][]
{\label{fig: seebeck_ac_phonon_full}\includegraphics[height=35mm,width=38mm]{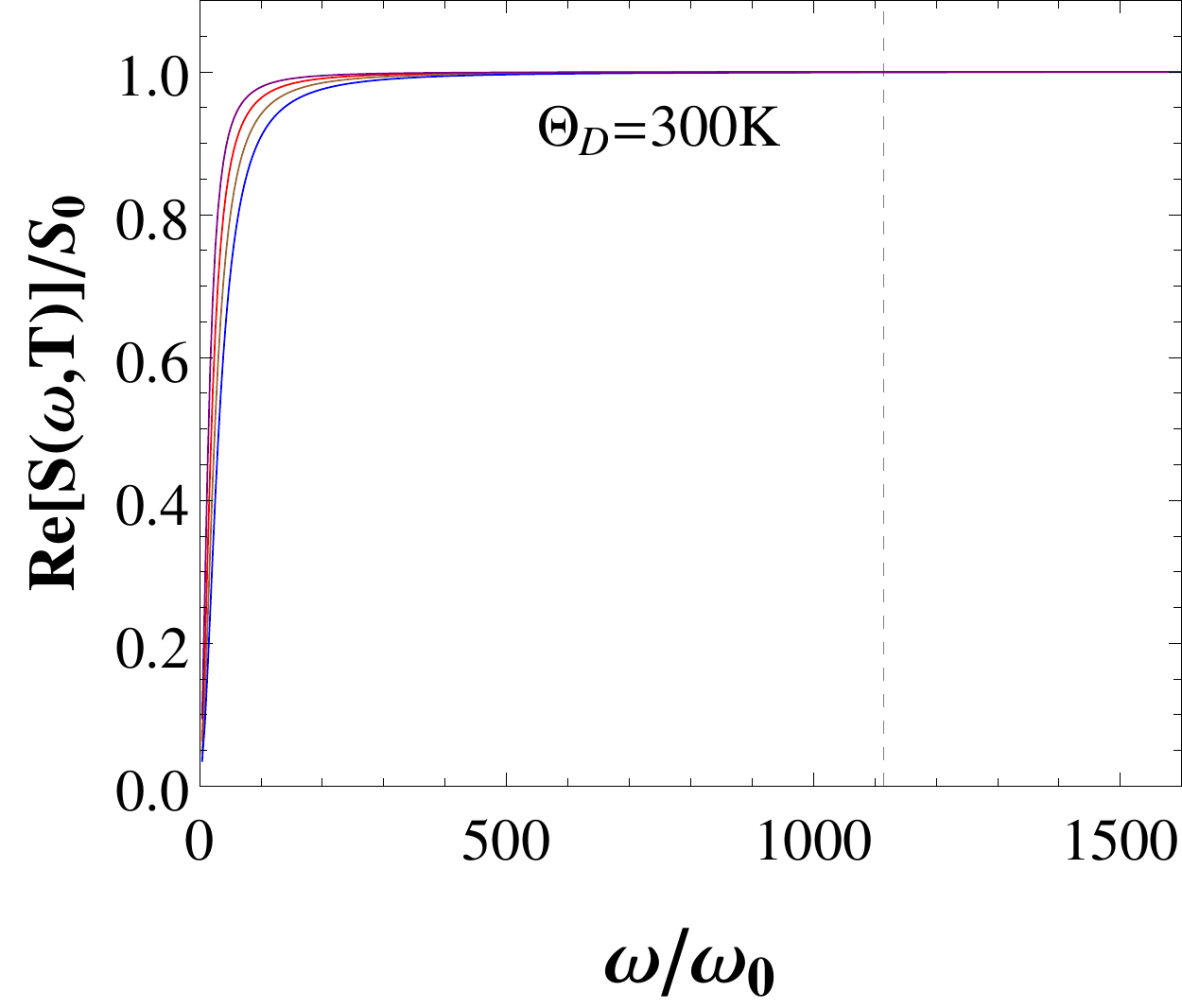}}
\hspace{0cm}
\subfigure[noonleline][]
{\label{fig: seebeck_ac_phonon_half}\includegraphics[height=35mm,width=40mm]{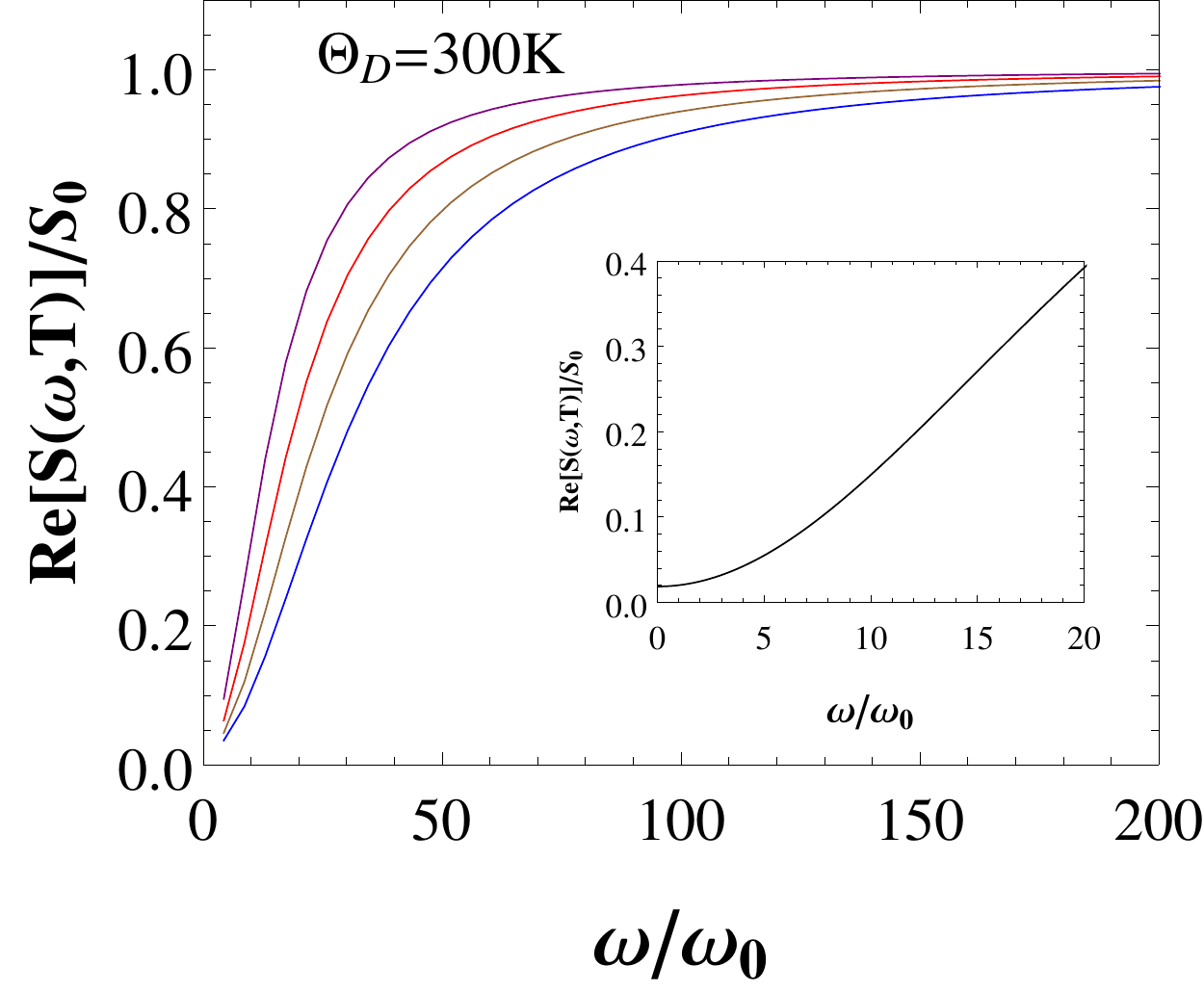}}
\caption{(a): Plot of finite frequency real part of the normalized Seebeck coefficient at different temperatures such as $200$(purple), $250$(red), $300$(brown), $350$(blue) and $375$K(magenta) at fixed Debye temperature $300$K. Here the dotted line corresponds to the Debye cuotff i.e. $\omega_{D}/\omega_{0}$, where $\omega_{0}$ is the constant scale parameter having dimensions of energy. (b): The low $\omega/\omega_{0}$ regime of fig.(a) is elaborated. Here $\omega_{0}$ is a scaling parameter.}
\label{fig: seebeck_ac_phonon}
\end{figure}
}
\newcommand{\figphononD}{
\begin{figure}[htb]
\centering
\hspace{0cm}
\subfigure[noonleline][]
{\label{fig: seebeck_dc_phonon_full}\includegraphics[height=35mm,width=40mm]{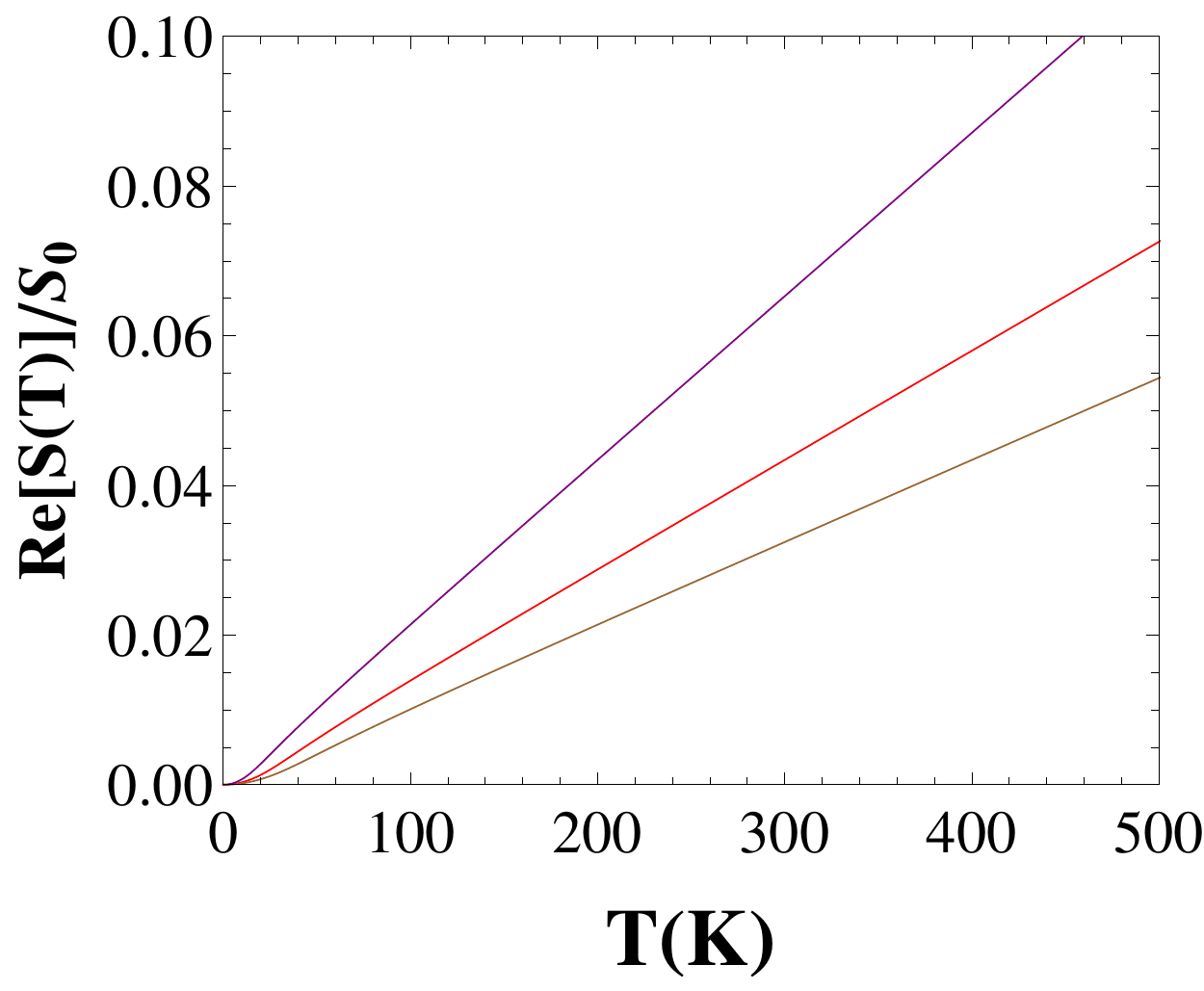}}
\hspace{0cm}
\subfigure[noonleline][]
{\label{fig: seebeck_dc_phonon_half}\includegraphics[height=35mm,width=40mm]{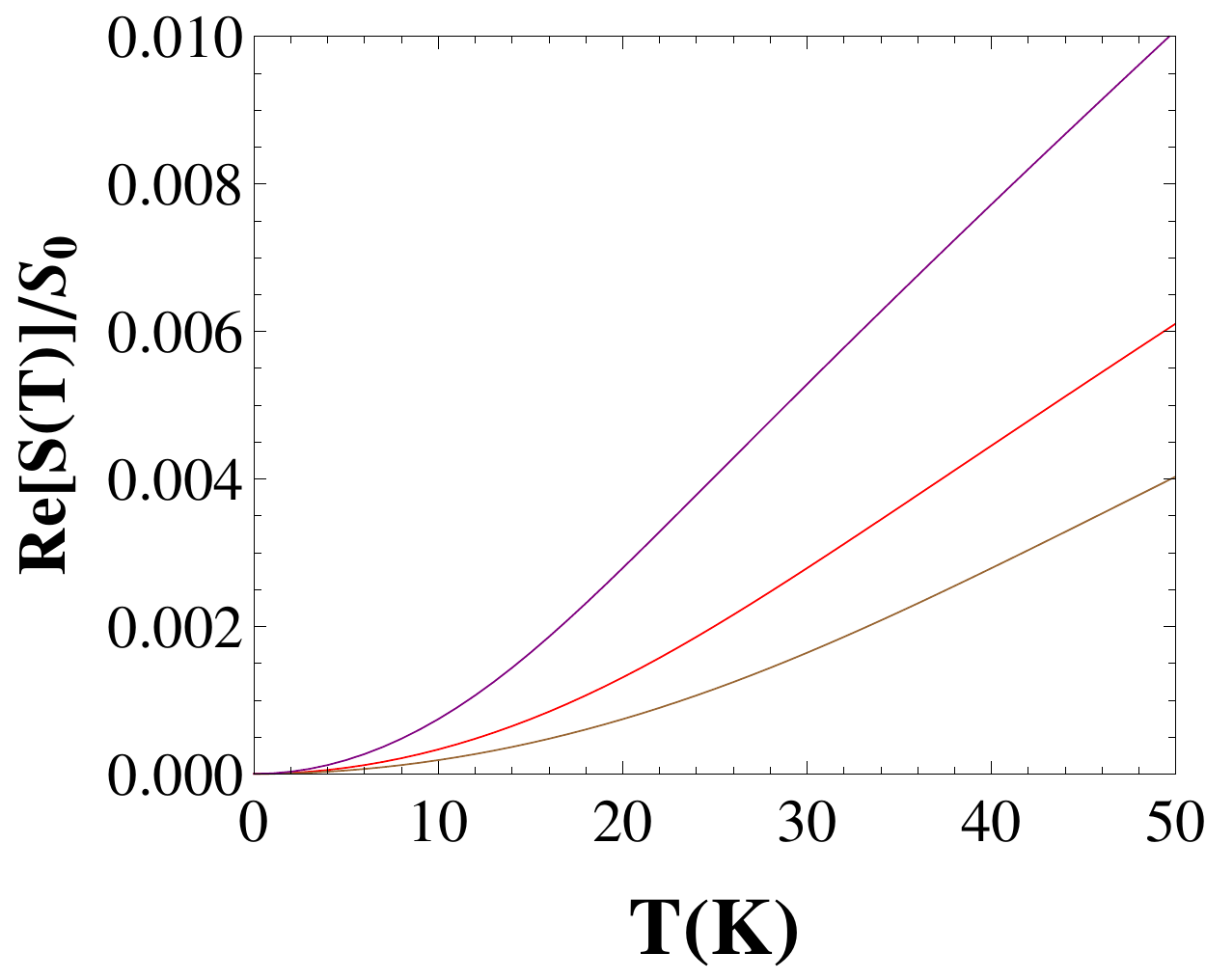}}
\caption{(a): Plot of the real part of the normalized Seebeck coefficient in zero frequency limit at different Debye temperatures such as $200$ (purple), $300$ (red) and $400$K (brown). (b): The low temperature regime of $\text{Re}[S(\omega, T)]/S_{0}$ of fig.(a) is elaborated.}
\label{fig: seebeck_dc_phonon}
\end{figure}
}
\newcommand{\figimpurityE}{
\begin{figure}[htb]
\centering
\hspace{0cm}
\subfigure[noonleline][]
{\label{fig: ac_impurity_full}\includegraphics[height=35mm,width=40mm]{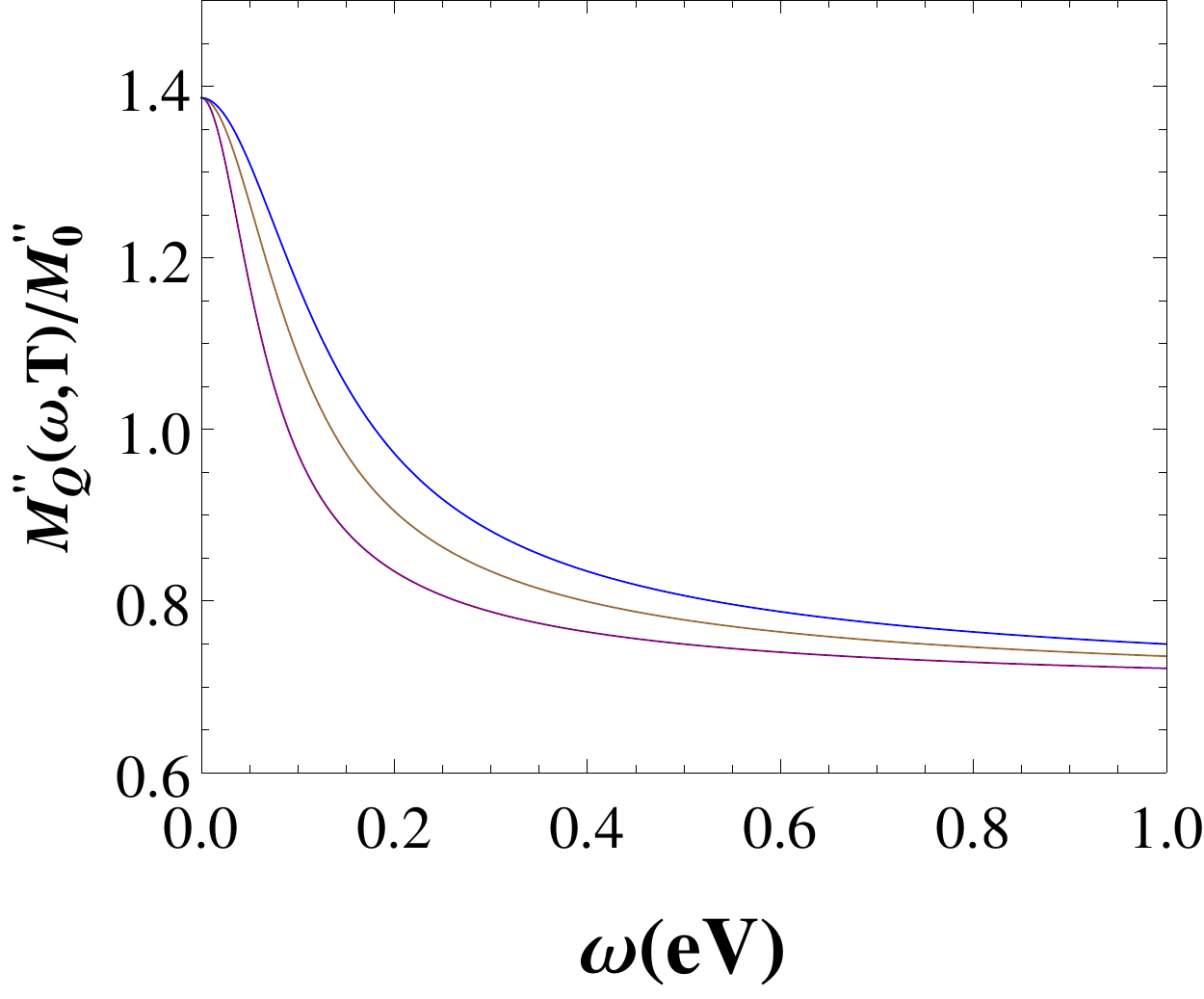}}
\hspace{0cm}
\subfigure[noonleline][]
{\label{fig: ac_impurity_half}\includegraphics[height=35mm,width=40mm]{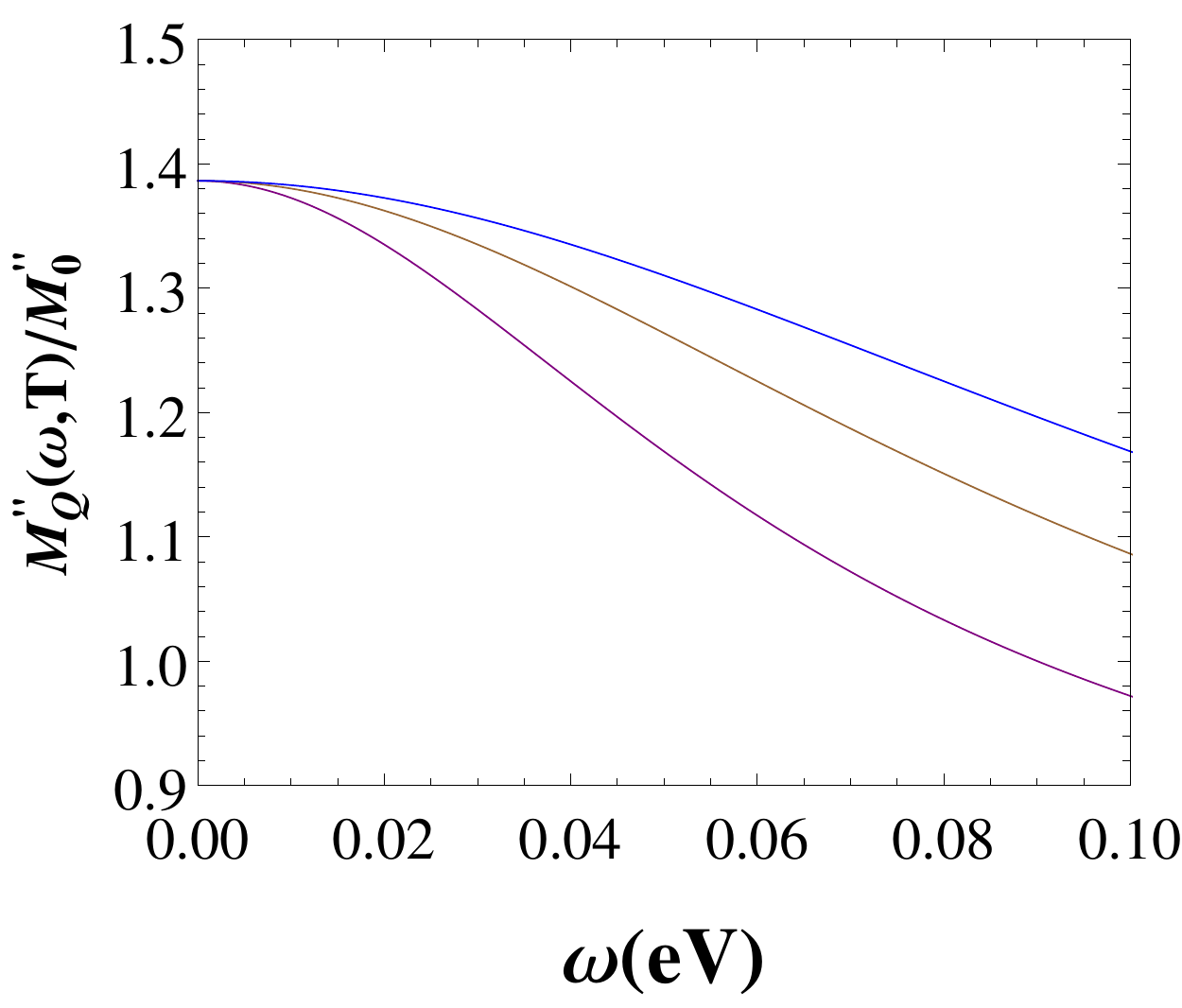}}
\caption{(a): Plot of the imaginary part of the thermoelectric memory function for impurity case at different temperatures such as $200$ K (0.017 eV, purple), $300$ K (0.026 eV, brown) and $400$ K (0.034 eV, blue). (b): The low frequency regime of $M''_{Q}(\omega, T)/M''_{0}$ of fig.(a) is elaborated.}
\label{fig: ac_impurity}
\end{figure}
}
\newcommand{\figimpurityF}{
\begin{figure}[htb]
\centering
\hspace{0cm}
\subfigure[noonleline][]
{\label{fig: ac_seebeck_impurity_full}\includegraphics[height=35mm,width=38mm]{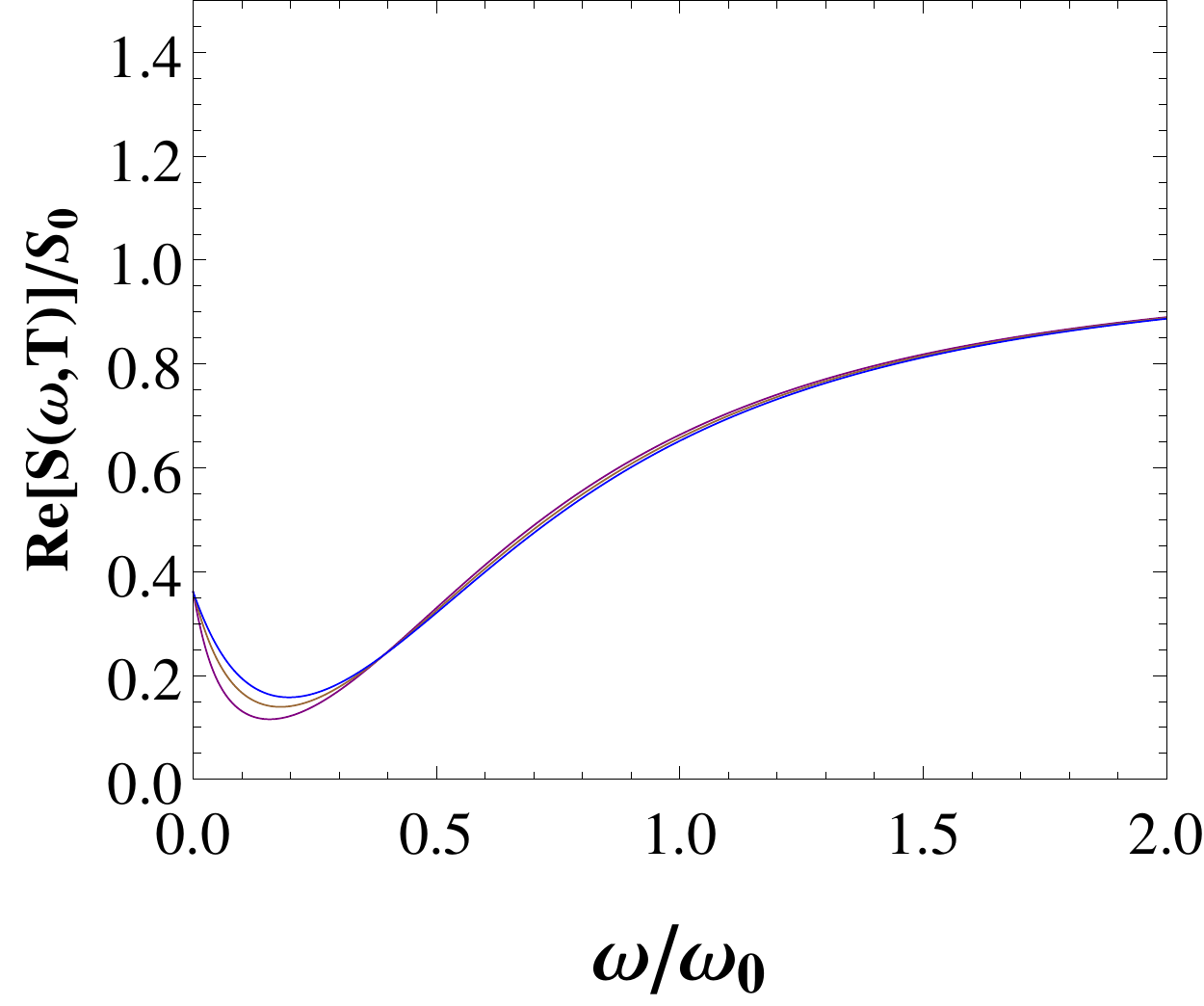}}
\hspace{0cm}
\subfigure[noonleline][]
{\label{fig: ac_seebeck_impurity_half}\includegraphics[height=35mm,width=40mm]{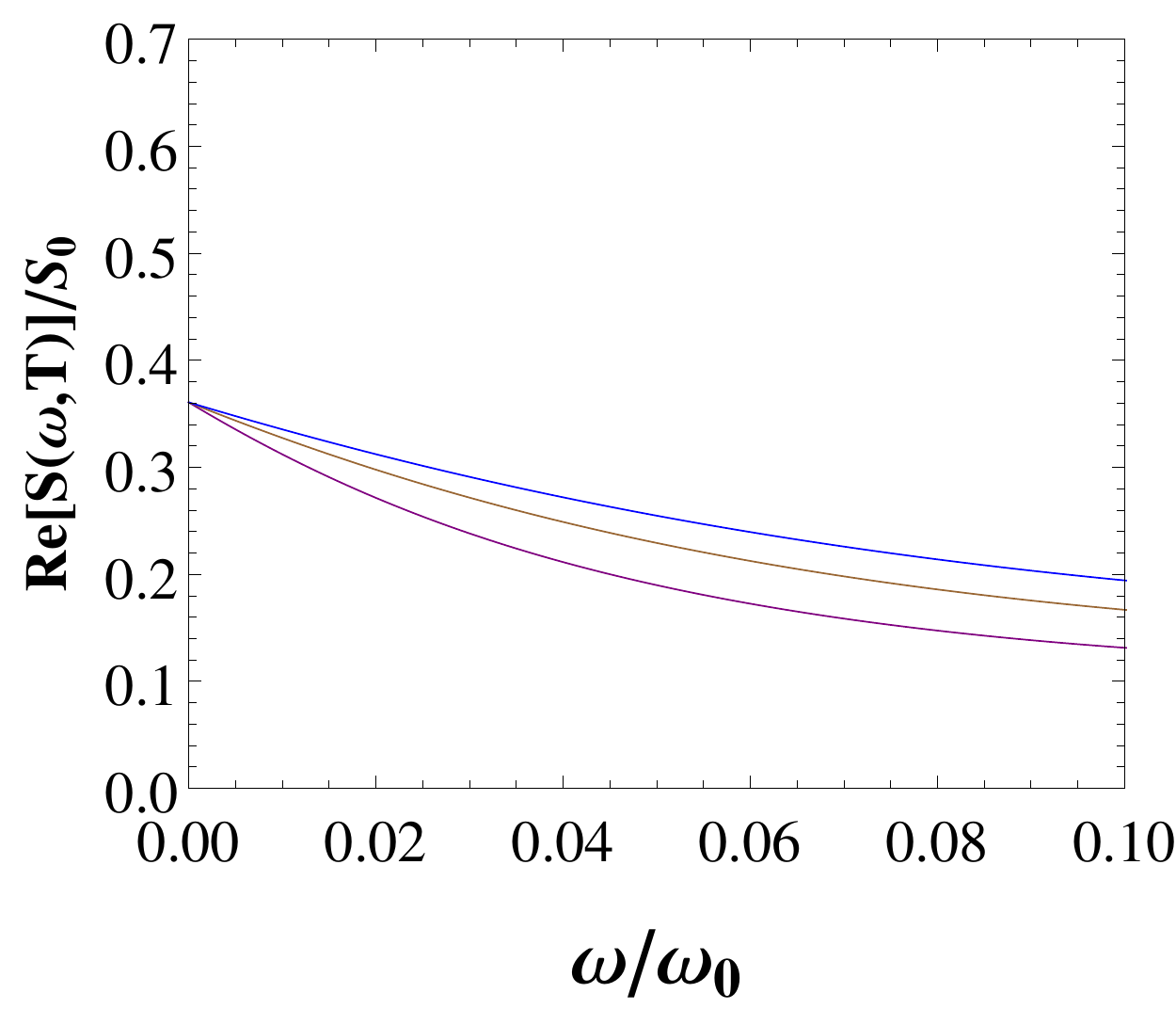}}
\caption{(a): Plot of the real part of the normalized Seebeck coefficient for impurity case at different temperatures such as $200$ K (0.017 eV, purple), $300$ K (0.026 eV, brown) and $400$ K (0.034 eV, blue). (b): The low frequency regime of $\text{Re}[S(\omega,T)/S_{0}$ of fig.(a) is elaborated.}
\label{fig: seebeck_ac_impurity}
\end{figure}
}
\begin{document}
\title{Finite frequency Seebeck coefficient of metals: A memory function approach}

\author{Pankaj Bhalla}
\email{pankajbhalla66@gmail.com}
\affiliation{Physical Research Laboratory, Navrangpura, Ahmedabad-380009 India.}
\affiliation{Indian Institute of Technology Gandhinagar-382424, India.}
\author{Pradeep Kumar}
\affiliation{Physical Research Laboratory, Navrangpura, Ahmedabad-380009 India.}
\affiliation{Indian Institute of Technology Gandhinagar-382424, India.}
\author{Nabyendu Das}
\affiliation{Physical Research Laboratory, Navrangpura, Ahmedabad-380009 India.}
\affiliation{Department of Physics, The LNM-Institute of Information Technology, Jaipur-302031, India.}
\author{Navinder Singh}
\affiliation{Physical Research Laboratory, Navrangpura, Ahmedabad-380009 India.}
\date{\today}
\begin{abstract}
We study the dynamical thermoelectric transport in metals subjected to the electron-impurity and the electron-phonon interactions using the memory function formalism. We introduce a generalized Drude form for the Seebeck coefficient in terms of thermoelectric memory function and calculate the later in various temperature and frequency limits. In the zero frequency and high temperature limit, we find that our results are consistent with the experimental findings and with the traditional Boltzmann equation approach. In the low temperature limit, we find that the Seebeck coefficient is quadratic in temperature. In the finite frequency regime, we report new results: In the electron-phonon interaction case, we find that the Seebeck coefficient shows frequency independent behavior both in the high frequency regime ($\omega \gg \omega_{D}$, where $\omega_{D}$ is the Debye frequency) and in the low frequency regime ($\omega \ll \omega_{D}$), whereas in the intermediate frequencies, it is a monotonically increasing function of frequency. In the case of the electron-impurity interaction,  first it decays and then after passing through a minimum it increases with the increase in frequency and saturates at high frequencies.
\end{abstract}
\maketitle

\section{Introduction}
\label{sec: intro}
In quest of modern technological advances to find highly efficient thermoelectric devices, the understanding of the thermoelectric transport is extremely important\cite{synder_08, zheng_08, bulusu_08}. From the industrial point of view, the thermoelectric devices require materials having large figure of merit ($ZT = S^2\sigma T / \kappa$), where $S$ is the Seebeck coefficient, $\sigma$ is the electrical conductivity, $\kappa$ is the thermal conductivity and $T$ is the temperature. Enormous efforts have been applied in order to increase the figure of merit so that the thermoelectric devices can be used for energy conversion processes\cite{rowe_book, shakouri_11, zebarjadi_12}. In the steady state, it can be increased by increasing the product of the electrical conductivity and square of the Seebeck coefficient i.e. $\sigma S^{2}$ or by decreasing the thermal conductivity $\kappa$. But in this pathway, there is a well known relation between the thermal conductivity and the electrical conductivity, known as Wiedemann-Franz law\cite{ashcroft_book}. The later makes it difficult to decrease $\kappa$ without the decrease of $\sigma$. Thus, an alternative approach known as dynamical approach is required to make this pathway easier\cite{ezzahri_14} which goes beyond the above mentioned restriction. We find that the Seebeck coefficient is higher at higher frequencies thus it leads to a greater figure of merit as the Wiedemann-Franz law is no more valid in the finite frequency case:
\bea
\frac{\kappa(\omega,T)}{T \sigma(\omega ,T)} &=& \frac{T M''_{QQ}(\omega,T)}{M''(\omega,T)} \left( \frac{\omega^2 + (M''(\omega,T))^{2}}{\omega^2 + (M''_{QQ}(\omega,T))^{2}}\right),
\eea
where $M''_{QQ}(\omega,T)$ and $M''(\omega,T)$ are thermal and electrical memory functions or known as scattering rates respectively. The right hand side of the above equation is not a constant. Thus Wiedemann-Franz restriction is not applicable.

Another importance of this study is due to the recent demand of the microelectronic and optoelectronic devices, working at several giga Hertz frequencies i.e. $GHz$ clock frequencies\cite{volz_01, koh_07}. The basic working principle of these devices involves various frequency dependent transport coefficients. Thus the quest of making these devices more efficient requires the understanding of the frequency and the temperature dependences of various transport quantities. So far, the dynamical nature of the electrical conductivity and the thermal conductivity have been studied in various recent works\cite{gotze_72, bhalla_16, bhalla_16a, nabyendu_16, bhalla_16b}. The study of the Seebeck coefficient is an important parameter to determine the figure of merit and was not studied previously in detail especially in the dynamical regime. Hence, in present work, we focus our study to investigate the dynamical behavior of the Seebeck coefficient.


On the other hand from the theoretical point of view, this problem is also of considerable interest. Firstly, these coefficients are generally calculated using the Bloch-Boltzmann approach under the relaxation time approximation (RTA)\cite{ziman_book}. It is found that the Seebeck coefficient, according to the well known Mott formula, shows the linear temperature behavior in the high temperature regime. In the present approach, we extend the traditional approaches based on the relaxation time approximation to the beyond relaxation time approximation which is achieved by the memory function formalism\cite{navinder_16}. 
Secondly, these coefficients also have been computed using the Kubo formalism where the two particle Green's function is generally written in terms of single particle Green's function\cite{mahan_book}. No such approximation is made in the present approach. 

With this motivation, we study the thermoelectric coefficients in both the static and finite frequency domain using the memory function formalism. The later, known as the projection operator technique is developed by Mori and Zwanzig and then extended by others\cite{forster_95, fulde_12, mori_65, zwanzig_61, zwanzig_61a, berne_66, berne_70, pires_88, harp_70, arfi_92, maldague_77, subir_14, lucas_15, subir_15, nabyendu_15}. This formalism is novel as it captures the frequency and temperature dependence of a dynamical response function in terms of the corresponding generalized scattering rate.

In the present work, we introduce for the first time and calculate the thermoelectric memory functions for the electron-impurity and the electron-phonon interactions and hence the Seebeck coefficient. Then, we compare our results in the case of zero frequency limit with the results predicted by the Bloch-Boltzmann approach and the experimental results. In the finite frequency case and in the case of the electron-phonon interaction, we find that the Seebeck coefficient increases with the increase of the frequency in the low frequency regime and then saturates to constant value in the high frequency regime ($\omega \gg \omega_{D}$). While in the case of electron-impurity interaction, it decreases with the increase of the frequency in low frequency regime ($\omega \ll \omega_{D}$) and then in the high frequency regime, it shows saturating behavior. 

This paper is organized as follows. In Sec.\ref{sec: thermoelectric coefficients}, we discuss the basic relations of the thermoelectric coefficients. Then in Sec.\ref{sec: memory}, first we briefly review the framework of the memory function formalism. Then, we introduce a model Hamiltonian and the thermoelectric memory functions. Here, we discuss the thermoelectric memory functions for the case of the electron-impurity and the electron-phonon interactions. We discuss these in the zero frequency and finite frequency limits. Later, we calculate the temperature and frequency variations of $S(z, T)$, where $z$ is the complex frequency. In Sec.\ref{sec: results}, we discuss our results and find that the results in the zero frequency limit are in accord with the experimental findings. Finally, we conclude in Sec\ref{sec: conclusion}.
\section{Thermoelectric coefficients}
\label{sec: thermoelectric coefficients}
In the linear response theory, the electric field and the temperature gradient are related to the electric current and the thermal current as follows\cite{mahan_book, kamran_book}.
\bea
\textbf{J} &=& \sigma \textbf{E} - \alpha \nabla T.
\label{eqn: electcurrent} \\
\textbf{J}_{Q} &=& \tilde{\alpha} \textbf{E} - \kappa \nabla T.
\label{eqn: thecurrent}
\eea
These equations tell that the generation of charge current and the flow of heat can be a consequence of either electric field or temperature gradient. Here $\sigma$ is the electrical conductivity, $\kappa$ is the thermal conductivity, $\alpha$ is the thermoelectric conductivity, and $\tilde{\alpha}$ is the electrothermal conductivity.\\
Consider that the system is electrically insulated. Thus, there is no electric current flow in the system i.e. $J = 0$. Therefore, Eq.(\ref{eqn: electcurrent}) can be written as
\bea
\frac{E}{\nabla T} &=& \frac{\alpha}{\sigma}.
\eea 
The Seebeck coefficient $S$ is defined as the electric field generated by a thermal gradient in the absence of electric current\cite{kamran_book}
\bea
S &=& -\frac{E}{\nabla T} = -\frac{\alpha}{\sigma}.
\label{eqn: seebeck}
\eea
Here the sign indicates the sign of the charge carriers.\\
The Peltier coefficient is defined as the flow of heat due to the electric current. According to the Kelvin relation, it can be expressed as\cite{noelle_book}
\bea
\Pi &=& ST.
\label{eqn: peltier}
\eea
Similarly, the Thomson coefficient which is related to the phenomenon of reversible heating or cooling in a current carrying material is defined as\cite{noelle_book}
\bea
\mu_{T} &=& T\frac{d S}{d T}
\label{eqn: thompson}
\eea
We see that all these coefficients are related (Eqs.(\ref{eqn: seebeck}) - (\ref{eqn: thompson})) and the calculation of the Seebeck coefficient is sufficient to understand the others. The former is the ratio of the thermoelectric conductivity and the electrical conductivity which are calculated in the following section.
\section{Memory Function Formalism}
\label{sec: memory}
According to the linear response theory, the correlation function is defined as\cite{kadanoff_63, kubo_57, zubarev_60}
\be
\chi_{AB}(z) = \langle \langle A ; B \rangle \rangle_{z} = -i\int_{0}^{\infty} e^{izt} \langle [A(t), B] \rangle dt.
\label{eqn: definitioncorrelation}
\ee
Here $A$ and $B$ are general operators corresponding to two different physical observables, $[A,B]$ is the commutator between operators $A$ and $B$ and the inner angular bracket in $\langle \langle \cdots \rangle \rangle$ corresponds to the thermodynamical average and the outer one for the Laplace transform at a complex frequency $z$.\\
This correlation function using equation of motion can be expressed as
\be
z\langle \langle A \vert B \rangle \rangle_{z} = \langle [A,B] \rangle + \langle \langle [A,H]; B \rangle \rangle_{z},
\label{eqn: eqmcorrelation}
\ee
where $H$ is the total Hamiltonian of the system. Presently we deal with thermoelectric responses. Thus we replace $A$ and $B$ by $J_{Q}$ and $J$ respectively, where $J_{Q}$ is the thermal current and $J$ is the electric current. Therefore
\be
z\langle \langle J_{Q} \vert J \rangle \rangle_{z} = \langle [J_{Q},J] \rangle + \langle \langle [J_{Q},H]; J \rangle \rangle_{z}.
\label{eqn: corrcommutator}
\ee
As the equal time commutator between the thermal current and the electric current vanishes. Thus, we left with the second term of Eq.(\ref{eqn: corrcommutator}). Hence on applying equation of motion on this term, the correlation function becomes
\bea \nonumber && 
\chi_{Q}(z,T) \\ &&=\frac{\langle \langle [J_{Q},H]; [J,H] \rangle \rangle_{z=0} - \langle \langle [J_{Q},H]; [J,H] \rangle \rangle_{z}}{z^2}.
\label{eqn: correlation}
\eea
Following the Refs.\cite{gotze_72, navinder_16}, the correlation function is related to the memory function as
\be
M_{Q}(z,T)=z\frac{\chi_{Q}(z,T)}{\chi_{Q}^{0}(T)-\chi_{Q}(z,T)},
\label{eqn: memorygeneral}
\ee
where $\chi_{Q}^{0}(T)$ is the static thermal current-electric current correlation function. On expanding Eq.(\ref{eqn: memorygeneral}), keeping the leading order term and using Eq.(\ref{eqn: correlation}), the memory function can be written as\cite{bhalla_16, bhalla_16b}
\bea \nonumber &&
M_{Q}(z,T)\\ &&=\frac{\langle \langle [J_{Q},H]; [J,H] \rangle \rangle_{z=0} - \langle \langle [J_{Q},H]; [J,H] \rangle \rangle_{z}}{z\chi_{Q}^{0}(T)}. 
\label{eqn: memory}
\eea
This is an thermoelectric memory function in terms of the thermal current and the electric current. This memory function relates to the thermoelectric response function as follows
\be
\alpha(z,T) = \frac{i}{T}  \frac{\chi_{Q}^{0}(T)}{z+M_{Q}(z,T)},
\label{eqn: thermal}
\ee
The above relation can be proved in a similar manner as done in the case of thermal response function\cite{bhalla_16b}. Following the Eqs.(\ref{eqn: memory}) and (\ref{eqn: thermal}), the thermoelectric memory function, hence the corresponding response function can be computed in various cases (e.g. electron-impurity, electron-phonon interactions) as done in the proceeding subsection.

\subsection{Model Hamiltonian}
\label{sec:modelhamiltonian}
Consider a system in which free electrons undergo scattering with impurities and phonons. In such a system, the total Hamiltonian is described as
\be
H = H_{0} + H_{\text{imp}} + H_{\text{ep}} + H_{\text{ph}}.
\label{eqn: hamiltonian}
\ee
The first part of the total Hamiltonian describes the free electrons and is given as
\be
H_{0} = \sum_{\textbf{k} \sigma} \epsilon_{\textbf{k}} c^{\dagger}_{\textbf{k} \sigma} c_{\textbf{k} \sigma},
\ee
where $\epsilon_{\textbf{k}}$ is the energy dispersion of free electrons, $c_{\textbf{k} \sigma}$($c^{\dagger}_{\textbf{k} \sigma}$) is annihilation(creation) operator having electronic momentum $\textbf{k}$ and spin $\sigma$. The second and the third part represent the electron-impurity and the electron-phonon interactions which are described as
\be
H_{\text{imp}} = N^{-1} \sum_{i} \sum_{\textbf{k} \textbf{k}' \sigma} \langle \textbf{k} \vert U^{i} \vert \textbf{k}^{\prime} \rangle c^{\dagger}_{\textbf{k} \sigma} c_{\textbf{k}' \sigma},
\label{eqn: impurityhamiltonian}
\ee
and
\be
H_{\text{ep}} = \sum_{\textbf{k} \textbf{k}' \sigma} \left[ D(\textbf{k}-\textbf{k}') c^{\dagger}_{\textbf{k} \sigma} c_{\textbf{k}' \sigma} b_{\textbf{k}-\textbf{k}'} + H.c. \right],
\label{phononhamiltonian}
\ee
respectively. Here $U^{i}$ refers to the impurity interaction strength, sum over $i$ index refers to the number of impurity sites and $N$ represents the number of lattice cells. The operator $b_{\textbf{q}}$($b_{\textbf{q}}^{\dagger}$) is the phonon annihilation(creation) operator having phonon momentum $\textbf{q}$ and $D(\textbf{q})$ is the electron-phonon matrix element. In case of metal, the later can be considered in the following form,\cite{ziman_book}
\be
D(\textbf{q}) = \frac{1}{\sqrt{2m_{i}N\omega_{q}}} q C(q).
\label{eqn: matrixelement}
\ee
The symbols used in the Eq.(\ref{eqn: matrixelement}): $m_{i}$ is the ionic mass, $\omega_{q}$ is the phonon frequency and $C(q)$ is a slowly varying function of $q$ which in case of metal is considered as $1/\rho_{F}$, where $\rho_{F}$ is the density of states at Fermi level \cite{ziman_book}. The last part of the Hamiltonian describes free phonons and can be written as
\be
H_{\text{ph}} = \sum_{q} \omega_{q} \left( b_{q}^{\dagger} b_{q} +\frac{1}{2} \right).
\label{freephononhamiltonian}
\ee
With this model Hamiltonian describing different perturbations due to the impurity and the phonon interactions with electrons, we calculate the thermoelectric memory function in next subsection.

\subsection{Thermoelectric Memory function}
\label{sec:thermoelectric_memoryfunction}
Before the calculation of the thermoelectric memory function $M_{Q}(z,T)$, we define the electric and thermal current in operator form as\cite{mahan_book}
\bea
J &=& \frac{1}{m} \sum_{\textbf{k}} \textbf{k}.\hat{n} c_{\textbf{k}}^{\dagger} c_{\textbf{k}}.
\label{eqn: electriccurrent}\\
J_{Q} &=& \frac{1}{m} \sum_{\textbf{k}} \textbf{k}.\hat{n} (\epsilon_{\textbf{k}} - \mu) c_{\textbf{k}}^{\dagger} c_{\textbf{k}}.
\label{eqn: thermalcurrent}
\eea
Here $\mu$ is the chemical potential, $m$ is the electron mass and $\hat{n}$ is the unit vector parallel to the direction of current.

With these definitions of currents, we calculate the thermoelectric memory function (defined in Eq.(\ref{eqn: memory})). 
\subsubsection{Electron-Impurity Interaction}
\label{sec:electronimpurityinteraction}
To compute $M_{Q}(z,T)$ for a system in which the total Hamiltonian is defined by $H = H_{0} + H_{\text{imp}}$ due to the presence of only electron-impurity interaction. First, we calculate the Laplace transform and thermal average of the inner product $\langle \langle [J_{Q},H] ; [J,H]\rangle \rangle_{z}$ which requires the commutation relations between the currents and the Hamiltonian. The commutator between the electric current and the Hamiltonian is given by $[J, H] = [J, H_{0}] + [J, H_{\text{imp}}]$. As the electric current and the unperturbed Hamiltonian commutes with each other, we have
\bea 
[J,H] &=& \frac{1}{m N} \sum_{i} \sum_{\textbf{k} \textbf{k}' \sigma}  \langle \textbf{k} \vert U^{i} \vert \textbf{k}' \rangle \left(\textbf{k} - \textbf{k}' \right).\hat{n}c^{\dagger}_{\textbf{k} \sigma} c_{\textbf{k}' \sigma}.
\label{eqn: electricimpuritycommutator}
\eea
Similarly, the commutator of thermal current and Hamiltonian is given by
\bea \nonumber
[J_{Q},H] &=& \frac{1}{m N} \sum_{i} \sum_{\textbf{k} \textbf{k}' \sigma}  \langle \textbf{k} \vert U^{i} \vert \textbf{k}' \rangle \\ 
&& \left(\textbf{k} (\epsilon_{\textbf{k}} - \mu) - \textbf{k}'(\epsilon_{\textbf{k}'}- \mu) \right).\hat{n}c^{\dagger}_{\textbf{k} \sigma} c_{\textbf{k}' \sigma}.
\label{eqn: thermalimpuritycommutator}
\eea
Using the above relations, the correlation function $\langle \langle [J_{Q},H] ; [J,H]\rangle \rangle_{z}$ becomes
\bea
\nonumber
= \frac{1}{m^2 N^2} \sum_{i j} \sum_{\textbf{k} \textbf{k}' \sigma} \sum_{\textbf{p} \textbf{p}' \tau} \langle \textbf{k} \vert U^{i} \vert \textbf{k}' \rangle \langle \textbf{p} \vert U^{j} \vert \textbf{p}' \rangle \\ \nonumber
\left(\textbf{k} - \textbf{k}' \right).\hat{n} \left(\textbf{p} (\epsilon_{\textbf{p}}- \mu) - \textbf{p}' (\epsilon_{\textbf{p}'} - \mu) \right).\hat{n} \\
 \langle \langle c^{\dagger}_{\textbf{k} \sigma} c_{\textbf{k}' \sigma} ; c^{\dagger}_{\textbf{p} \tau} c_{\textbf{p}' \tau} \rangle \rangle_{z}.
\label{eqn: doubleannularbracket}
\eea
Considering the case of $i=j$ (case of dilute impurity in which interaction terms $i \ne j$ are neglected), performing ensemble average and integrating over time, Eq.(\ref{eqn: doubleannularbracket}) reduces to
\bea \nonumber
&=& \frac{2 N_{\text{imp}}}{m^2 N^2} \sum_{\textbf{k} \textbf{k}'}  \vert \langle \textbf{k} \vert U \vert \textbf{k}' \rangle \vert^{2}\left(\textbf{k} - \textbf{k}' \right).\hat{n} \\ 
&& \left(\textbf{k} (\epsilon_{\textbf{k}}- \mu) - \textbf{k}' (\epsilon_{\textbf{k}'} - \mu) \right).\hat{n}  \times \frac{f_{\textbf{k}}-f_{\textbf{k}'}}{z+\epsilon_{\textbf{k}}-\epsilon_{\textbf{k}'}}.
\label{eqn: finaldoubleannularbracket}
\eea
Here $f_{\textbf{k}} = \frac{1}{e{\frac{\epsilon_{\textbf{k}} - \mu}{T}}+1}$ is the Fermi distribution function.
Substituting Eq.(\ref{eqn: finaldoubleannularbracket}) in the thermoelectric memory function (Eq.(\ref{eqn: memory})) and performing the analytic continuation using $z \rightarrow \omega + i\eta$, $\eta \rightarrow 0^{+}$, the imaginary part of the thermoelectric memory function is expressed as
\bea \nonumber
M''_{Q}(\omega, T) &=& \frac{2\pi N_{\text{imp}}}{\chi_{Q}^{0}(T) m^2 N^2} \sum_{\textbf{k} \textbf{k}'} \vert \langle \textbf{k} \vert U \vert \textbf{k}' \rangle \vert ^{2} \left( \textbf{k} - \textbf{k}' \right).\hat{n}\\ \nonumber
&&\left( \textbf{k} (\epsilon_{\textbf{k}} - \mu) - \textbf{k}'(\epsilon_{\textbf{k}'} - \mu) \right).\hat{n} \\ 
&& \frac{f_{\textbf{k}}-f_{\textbf{k}'}}{\omega} \delta(\omega+\epsilon_{\textbf{k}}-\epsilon_{\textbf{k}'}).
\label{eqn: memorythermalimpurity2}
\eea
To simplify the Eq.(\ref{eqn: memorythermalimpurity2}), it is assumed that the system has cubic symmetry. Thus using the laws of conservation of energy and conservation of momentum, the part of above equation can be written as
\bea \nonumber
\left( \textbf{k} - \textbf{k}' \right).\hat{n} \left( \textbf{k} (\epsilon_{\textbf{k}} - \mu) - \textbf{k}'(\epsilon_{\textbf{k}'} - \mu) \right).\hat{n}\\
= \frac{1}{3}\left((\epsilon_{\textbf{k}} - \mu) k^2 +(\epsilon_{\textbf{k}'} - \mu) k'^2 \right).
\label{eqn: smallpart}
\eea
Using the Eq.(\ref{eqn: smallpart}) and considering the momentum independent character of the impurity strength $U$, the Eq.(\ref{eqn: memorythermalimpurity2}) in the integral form reduces to 
\bea \nonumber
M''_{Q}(\omega, T) &=& \frac{N_{\text{imp}} U^2 k_{F}^4}{6 \pi^3 \chi_{Q}^{0}(T)} \int_{0}^{\infty} d\epsilon \left( 2(\epsilon_{\textbf{k}} - \mu) + \omega \right) \\
&&\times \frac{f(\epsilon_{\textbf{k}})-f(\epsilon_{\textbf{k}}+\omega)}{\omega}. 
\label{eqn: memorybeforedimensionless}
\eea
Here we consider the scattering events occurring only near the Fermi surface. Thus the magnitude of electron momentum $\textbf{k}$ and $\textbf{k}'$ are approximately equal to $k_{F}$, the Fermi wave vector.\\
Defining the new dimensionless variables $\frac{\epsilon_{\textbf{k}} -\mu}{T} = \eta$ and $\frac{\omega}{T} = x$ and substituting it in Eq.(\ref{eqn: memorybeforedimensionless}) we have
\bea \nonumber
M''_{Q}(\omega, T) &=& \frac{N_{\text{imp}} U^2 k_{F}^4 T}{6 \pi^3 \chi_{Q}^{0}(T)} \int_{0}^{\infty} d\eta \frac{2\eta + x}{x} \\
&& \left( \frac{1}{e^{\eta}+1} - \frac{1}{e^{\eta + x}+1} \right).
\label{eqn:memoryimpurity_final}
\eea
This is an expression for the imaginary part of the thermoelectric memory function in the presence of electron-impurity interaction. Its behavior can be discussed in different frequency and temperature regimes as follows.\\
\textbf{Case-I The zero frequency limit i.e. $\omega \rightarrow 0$:}\\
In this limit, Eq.(\ref{eqn:memoryimpurity_final}) can be written as
\bea
M_{Q}''(T) = \frac{N_{\text{imp}}}{3 \pi^3} \frac{U^2 k_{F}^{4} T}{\chi_{Q}^{0}(T)} \int_{0}^{\infty} d\eta \frac{\eta e^{\eta}}{(e^{\eta}+1)^2}.
\label{eqn:dcthermomem}
\eea
This expression shows that the imaginary part of the zero frequency thermoelectric memory function varies with temperature as $T/\chi_{Q}^{0}(T)$. As proved in the appendix \ref{app: static_derivation} that the static thermoelectric correlation function depends linearly on the temperature. Thus, in the zero frequency limit, $M_{Q}''(T)$ is independent of the temperature. Using this temperature variation of $M_{Q}''(T)$, the thermoelectric response function (Eq.(\ref{eqn: thermal})) in the zero frequency limit becomes
\bea \nonumber
\alpha(T) &=& \frac{1}{T} \frac{\chi_{Q}^{0}(T)}{M_{Q}''(T)}. 
\label{eqn:dc_ETC}
\eea
Thus we concludes that the thermoelectric conductivity shows temperature independent behavior in the case of electron-impurity.\\
\textbf{Case-II The finite frequency regime}\\
In the high frequency limit i.e. $\omega \gg T$, the Eq.(\ref{eqn:memoryimpurity_final}) reduces to
\bea \nonumber
M''_{Q}(\omega, T) &\approx& \frac{N_{\text{imp}} U^2 k_{F}^4 T}{6 \pi^3 \chi_{Q}^{0}(T)} \left(\frac{1-2e^{-\omega/T}}{\omega/T}  + e^{-\omega/T}+\log2  \right).\\
&\approx& \frac{N_{\text{imp}} U^2 k_{F}^4}{6 \pi^3 }\frac{T}{\chi_{Q}^{0}(T)} \log 2.
\eea
In the opposite case, when $\omega \ll T$, the imaginary part of the thermoelectric memory function (Eq.(\ref{eqn:memoryimpurity_final})) with the leading order term becomes
\bea
M''_{Q}(\omega, T) &\approx& \frac{N_{\text{imp}} U^2 k_{F}^4 }{18 \pi^3 } \frac{T}{\chi_{Q}^{0}(T) }.
\eea
The detailed analysis of these asymptotic results is summarized in Table \ref{tab: thermoelectric_scattering} and is discussed in later sections.
\subsubsection{Electron-Phonon Interaction}
\label{sec:electronphononinteraction}
In the presence of only electron-phonon interaction in a system, the total Hamiltonian is described by $H = H_{0} + H_{\text{ep}} + H_{\text{ph}}$. With this Hamiltonian, the thermoelectric memory function can be calculated using the commutation relations of current and $H$ in a similar way as done in the case of impurity interaction. In this electron-phonon interaction case, the commutation relations are defined as follows
\bea \nonumber
[J,H] &=& \frac{1}{m} \sum_{\textbf{k}, \textbf{k}', \sigma} \left( \textbf{k} - \textbf{k}' \right).\hat{n} \\
&& \left( D(\textbf{k}-\textbf{k}') c_{\textbf{k} \sigma}^{\dagger} c_{\textbf{k}' \sigma} b_{\textbf{k}-\textbf{k}'} - H.c. \right). \\  \nonumber
\left[J_{Q},H\right] &=& \frac{1}{m} \sum_{\textbf{k}, \textbf{k}', \sigma} \left( \textbf{k}(\epsilon_{\textbf{k}}-\mu) - \textbf{k}'(\epsilon_{\textbf{k}'}-\mu) \right).\hat{n} \\ \nonumber
&& \left( D(\textbf{k}-\textbf{k}') c_{\textbf{k} \sigma}^{\dagger} c_{\textbf{k}' \sigma} b_{\textbf{k}-\textbf{k}'} - H.c. \right).\\
\eea
Using these relations and after simplifications, $\langle \langle [J_{Q},H] ; [J,H]\rangle \rangle_{z}$ can be written as
\bea \nonumber
&=& \frac{2}{m^2} \sum_{\textbf{k} \textbf{k}'} \left( \textbf{k} - \textbf{k}'\right).\hat{n} \left( \textbf{k} (\epsilon_{\textbf{k}} - \mu) - \textbf{k}' (\epsilon_{\textbf{k}'} - \mu) \right).\hat{n}\\ \nonumber
&& \vert D(\textbf{k}-\textbf{k}') \vert^{2} f_{\textbf{k}'}(1-f_{\textbf{k}}) n_{\textbf{k}-\textbf{k}'}\left(e^{\beta(\epsilon_{\textbf{k}'}-\epsilon_{\textbf{k}} + \omega_{\textbf{k}-\textbf{k}'})} -1\right)
\\ \nonumber
&& \left\lbrace \frac{1}{z-\epsilon_{\textbf{k}'}+\epsilon_{\textbf{k}} - \omega_{\textbf{k}-\textbf{k}'}} - \frac{1}{z+\epsilon_{\textbf{k}'}-\epsilon_{\textbf{k}} + \omega_{\textbf{k}-\textbf{k}'}} \right\rbrace. \\
\label{eqn: finaldoubleannularbracketphonon}
\eea
Here $n_{\textbf{k}-\textbf{k}'} = \frac{1}{e^{\beta\omega_{\textbf{k}-\textbf{k}'}}-1}$ is a Boson distribution function at a temperature $T=1/\beta$.\\
Putting Eq.(\ref{eqn: finaldoubleannularbracketphonon}) in the thermoelectric memory function (Eq.(\ref{eqn: memory})), the imaginary part of the thermoelectric memory function after performing analytic continuation $z \rightarrow \omega + i\eta$, $\eta \rightarrow 0^{+}$ can be expressed as
\bea \nonumber
M''_{Q}(\omega, T) &=& \frac{2\pi}{\chi_{Q}^{0}(T) m^2} \sum_{\textbf{k} \textbf{k}'} \left(\textbf{k}-\textbf{k}' \right).\hat{n}\\ \nonumber
&& \left( \textbf{k} (\epsilon_{\textbf{k}} - \mu) - \textbf{k}'(\epsilon_{\textbf{k}'} - \mu) \right).\hat{n} \\ \nonumber
&& \vert D(\textbf{k}-\textbf{k}') \vert^{2} (1-f_{\textbf{k}})f_{\textbf{k}'} n_{\textbf{k}-\textbf{k}'} \\ \nonumber
&& \left\lbrace \frac{e^{\omega/T}-1}{\omega} \delta(\epsilon_{\textbf{k}}-\epsilon_{\textbf{k}'} - \omega_{\textbf{k}-\textbf{k}'} + \omega) \right. \\
&& \left. + (\text{terms with $\omega \rightarrow -\omega$}) \right\rbrace.
\label{eqn: memoryphonon2}
\eea
To simplify the above expression, we use the law of conservation of energy $\epsilon_{\textbf{k}} = \epsilon_{\textbf{k}'} - \omega_{q}$ and  the law of conservation of momentum $\textbf{k}'-\textbf{k} = \textbf{q}$. Thus,
\bea \nonumber
&&\left(\textbf{k}-\textbf{k}' \right).\hat{n}
\left( \textbf{k} (\epsilon_{\textbf{k}} - \mu) - \textbf{k}'(\epsilon_{\textbf{k}'} - \mu) \right).\hat{n}\\
&& = \textbf{q}.\hat{n} \left( \textbf{q}.\hat{n} (\epsilon_{\textbf{k}} - \mu) + \textbf{k}'.\hat{n}\omega_{\textbf{q}} \right).
\eea
For simplifications, we consider that the system has cubic symmetry, thus the averaging over all the directions reduces the above expression as $\frac{1}{3}\left(q^2 (\epsilon_{\textbf{k}} - \mu) - k'^{2} \omega_{\textbf{q}}\right)$. Using this relation and converting the summations into integrals along with introducing the new dimensionless variables $ \frac{\epsilon_{\textbf{k}}-\mu}{T} = \eta$, $\frac{\omega}{T} = x$ and $\frac{\omega_{q}}{T} = y$, the imaginary part of the thermoelectric memory function (Eq.(\ref{eqn: memoryphonon2})) becomes
\bea \nonumber
M''_{Q}(\omega, T) &=& \frac{N^2 T^3}{12\pi^3 \chi_{Q}^{0}(T)} \left(\frac{q_{D}}{\Theta_{D}}\right)^2 \int_{0}^{\infty} d\eta \int_{0}^{\Theta_{D}/T} dy  \\ \nonumber
&& \vert D(y) \vert^{2}\frac{y}{e^{y}-1} \frac{1}{e^{-\eta} + 1} \left\lbrace \eta \left(\frac{q_{D}T}{\Theta_{D}}\right)^2 y^2 - k_{F}^2 y\right\rbrace \\
&& \left[ \frac{e^{x} -1}{x(e^{\eta - y + x}+1)} + (\text{terms with $\omega \rightarrow -\omega$}) \right].
\eea
Substituting the phonon matrix element (Eq.\ref{eqn: matrixelement}) and solving the integral over $\eta$, we obtain
\bea \nonumber
M''_{Q}(\omega, T) &=& \frac{N T^4}{48\pi^3 m_{i} \rho_{F}^2 \chi_{Q}^{0}(T)} \left(\frac{q_{D}}{\Theta_{D}}\right)^4 \int_{0}^{\Theta_{D}/T} dy  \\ \nonumber
&& \frac{y^2}{e^{y}-1} \left\lbrace \frac{x-y}{e^{x-y} -1} \frac{e^x -1}{x} \right. \\ \nonumber
&& \left. \left((x-y) \left(\frac{q_{D}T}{\Theta_{D}}\right)^2 y^2 - 2 k_{F}^2 y\right)\right. \\
&& \left. + (\text{terms with $\omega \rightarrow -\omega$}) \right\rbrace.
\label{eqn: TE_MF_phonon}
\eea
This is the final expression of the imaginary part of the thermoelectric memory function in the case of the electron-phonon interaction. In different limits of frequency and temperature, it is discussed as follows.\\
\textbf{Case-I The zero frequency limit:}\\
In this limit i.e. $\omega \rightarrow 0$, the magnitude of the imaginary part of the thermoelectric memory function (Eq.\ref{eqn: TE_MF_phonon}) reduces as
\bea \nonumber
M''_{Q}(T) &=& \frac{N T^4}{24\pi^3 m_{i} \rho_{F}^2 \chi_{Q}^{0}(T)} \left(\frac{q_{D}}{\Theta_{D}}\right)^4 \int_{0}^{\Theta_{D}/T} dy  \\ 
&& \frac{y^4 e^{y}}{(e^{y}-1)^2}\left(\left(\frac{q_{D}T}{\Theta_{D}}\right)^2 y^2 + 2 k_{F}^2 \right).
\label{eqn: TE_MF_phonon_DC}
\eea
This expression can be further discussed in two subcases for high and low temperature regimes.\\
In the high temperature regime i.e. $T \gg \Theta_{D}$, the first term within a bracket of Eq.(\ref{eqn: TE_MF_phonon_DC}) i.e. $\left(\frac{q_{D}T}{\Theta_{D}}\right)^2 y^2$ gives more contribution to $M''_{Q}(T)$. Thus, the later becomes
\bea \nonumber
M''_{Q}(T) &\approx & \frac{N T^6}{24\pi^3 m_{i} \rho_{F}^2 \chi_{Q}^{0}(T)} \left(\frac{q_{D}}{\Theta_{D}}\right)^6 \int_{0}^{\Theta_{D}/T} dy   \frac{y^6 e^{y}}{(e^{y}-1)^2}. \\
&\approx & \frac{N}{24\pi^3 m_{i} \rho_{F}^2} \left(\frac{q_{D}}{\Theta_{D}}\right)^6 \frac{T}{\chi_{Q}^{0}(T)}.
\label{eqn: TE_MF_phonon_DC_high}
\eea
In the opposite case when $T \ll \Theta_{D}$, Eq.(\ref{eqn: TE_MF_phonon_DC}) becomes
\bea 
M''_{Q}(T) & \approx & \frac{2 N k_{F}^2 T^4}{\pi^3 m_{i} \rho_{F}^2 \chi_{Q}^{0}(T)} \left(\frac{q_{D}}{\Theta_{D}}\right)^4 \frac{T^4}{\chi_{Q}^{0}(T)}.
\label{eqn: TE_MF_phonon_DC_low}
\eea
Thus from these above expressions we find that the imaginary part of the thermoelectric memory function in the zero frequency limit is proportional to the $T/\chi_{Q}^{0}(T)$  in the high and $T^4/\chi_{Q}^{0}(T)$ in the low temperature regimes. The static thermoelectric correlation $\chi_{Q}^{0}(T)$ varies linearly with the temperature (as given in appendix \ref{app: static_derivation}). Thus, $M''_{Q}(T)$ varies as $T^3$ in the low temperature regime and becomes saturated in the high temperature regime. Substituting this in Eq.(\ref{eqn: thermal}), we find that the thermoelectric response function in the zero frequency limit shows temperature dependence as
\bea
\alpha(T) &=& \frac{1}{T} \frac{\chi_{Q}^{0}(T)}{M''_{Q}(T)}.
\eea
Hence, it varies as $T^{-3}$ in the low temperature regime and becomes temperature independent in the high temperature regime i.e.
\bea
\alpha(T) &\approx& \begin{cases}
  T^{-3},& T\ll \Theta_{D}\\
  \textnormal{constant},& T\gg \Theta_{D}
\end{cases}
\eea
\textbf{Case-II The finite frequency regime:}\\
In the high frequency limit i.e. when the frequency is more than the Debye frequency ($\omega \gg \omega_{D}$), the imaginary part of the thermoelectric memory function (Eq.(\ref{eqn: TE_MF_phonon})) can be written as
\bea \nonumber
M''_{Q}(\omega, T) &\approx&  \frac{N k_{F}^2}{12 \pi^3 m_{i} \rho_{F}^2 } \left( \frac{q_{D}}{\Theta_{D}} \right)^4 \frac{T^4}{\chi_{Q}^{0}(T)} \int_{0}^{\Theta_{D}/T} dy \frac{y^3}{e^y -1}.\\ \nonumber
&\approx& \frac{N k_{F}^2}{12 \pi^3 m_{i} \rho_{F}^2 } \left( \frac{q_{D}}{\Theta_{D}} \right)^4 \\ \nonumber
&& \times \frac{T^4}{\chi_{Q}^{0}(T)}
\begin{cases}
  \frac{\pi^4}{15},& T\ll \Theta_{D} \\
  \frac{1}{3}\left(\frac{\Theta_{D}}{T} \right)^3,&  T\gg \Theta_{D}.
\end{cases} \\
\eea
On substituting the temperature variation of $\chi_{Q}^{0}(T)$, the thermoelectric memory function shows temperature dependencies as
\bea
M''_{Q}(\omega, T) &\approx& \begin{cases}
  T^{3},& T \ll \Theta_{D}\\
  \textnormal{constant},& T\gg \Theta_{D}.
  \end{cases}
\eea
This implies that the imaginary part of the thermal memory function in case of the electron-phonon interaction becomes independent of frequency at high frequency regime. In this regime, it shows $T^3$ behavior in the low temperature regime and temperature independent behavior in the high temperature regime.\\
Similarly in the low frequency regime i.e. $\omega \ll \omega_{D}$, $M''_{Q}(\omega,T)$ (Eq.(\ref{eqn: TE_MF_phonon})) in the leading order is given by
\bea \nonumber
M''_{Q}(\omega, T) &\approx& \frac{N}{24 \pi^3 m_{i} \rho_{F}^2} \left( \frac{q_{D}}{\Theta_{D}} \right)^4 \frac{T^5}{\chi_{Q}^{0}(T)} \frac{\sinh \left(\omega/T\right)}{\omega}\\ \nonumber
&& \times \int_{0}^{\Theta_{D}/T} dy \frac{y^4 e^y}{(e^y-1)^2} \left\lbrace \left( \frac{q_{D} T}{\Theta_{D}} \right)^2 y^2 + 2k_{F}^2 \right\rbrace. \\ 
\eea
Now in the limit $\omega \gg T$,
\bea 
M''_{Q}(\omega, T) &\approx& \frac{N \pi k_{F}^{2}}{45 m_{i} \rho_{F}^2} \left( \frac{q_{D}}{\Theta_{D}} \right)^4 \frac{e^{\omega/T}}{\omega} \frac{T^5}{\chi_{Q}^{0}(T)}.  
\eea
In the opposite case, i.e. $\omega \ll T$,
\bea \nonumber
M''_{Q}(\omega, T) &\approx& \frac{N}{24 \pi^3 m_{i} \rho_{F}^2} \left( \frac{q_{D}}{\Theta_{D}} \right)^4  \\ \nonumber
&& \times \frac{T^4}{\chi_{Q}^{0}(T)} 
\begin{cases}
  \frac{8\pi^4}{15} k_{F}^2,& \textnormal{at $T\ll \Theta_{D}$} \\
  \frac{1}{5}q_{D}^2\left(\frac{\Theta_{D}}{T} \right)^3,&\textnormal{at $T\gg \Theta_{D}$}
\end{cases}. \\ 
\eea
This concludes that the finite frequency imaginary part of the thermoelectric memory function shows frequency dependence of the form $e^{\omega/T}/\omega$ in the regime where the frequency is more than the temperature and becomes frequency independent in the opposite case i.e. $\omega \ll T$. There is also different temperature dependences within both the former regimes depending on whether the temperature is greater or lesser than the Debye temperature. The details of these asymptotic results is discussed in later sections and are collected in Table (\ref{tab: thermoelectric_scattering}).

\begin{widetext}
\begin{table}[b]
\caption{The thermoelectric scattering rate due to the electron-impurity and the electron-phonon interactions in different frequency and temperature domains.}
\begin{center}
  \begin{tabular}{|c| c| c| c| c|} 
 \hline
 \multicolumn{3}{|c|}{\textbf{Electron-impurity interaction}}\\
 \hline
 $\omega = 0$ & $\omega \gg T$ & $\omega \ll T$\\ 
  \hline 
  $1/\tau_{\text{te}} \sim T^{0}$ & \begin{tabular}{c} 
 $1/\tau_{\text{te}} \sim \log2 + \frac{T}{\omega}$ 
 \end{tabular} 
 & $1/\tau_{\text{te}} \sim T^{0}$ \\
  \hline
   \multicolumn{3}{|c|}{\textbf{Electron-Phonon interaction}}\\
 \hline
 $\omega = 0$ & $\omega \gg \omega_{D}$ & $\omega \ll \omega_{D}$ \\ 
 \hline
 \begin{tabular}{c | c} 
  $T \gg \Theta_{D}$ & $T \ll \Theta_{D}$\\ 
 \hline
 $1/\tau_{\text{te}} \sim T^{0}$ &$1/\tau_{\text{te}} \sim T^{3} $ \\ 
 \end{tabular} & \begin{tabular}{c| c} 
  $\omega \gg T$ & $\omega \ll T$\\  
 \hline
 \begin{tabular}{c| c} 
 $T \gg \Theta_{D}$ & $T \ll \Theta_{D}$\\ 
 \hline
 $1/\tau_{\text{te}} \sim T^{0}$ &$1/\tau_{\text{te}} \sim T^{3} $ \\ 
 \end{tabular} &\begin{tabular}{c} 
  $T \gg \Theta_{D}$ \\ 
 \hline
 $1/\tau_{\text{te}} \sim T^{0}$ \\ 
 \end{tabular} \\ 
\end{tabular} & \begin{tabular}{c| c} 
  $\omega \gg T$ & $\omega \ll T$\\ 
 \hline
 \begin{tabular}{c} 
  $T \ll \Theta_{D}$ \\ 
 \hline
 $1/\tau_{\text{te}} \sim T^{4} \frac{e^{\omega/T}}{\omega}$ \\
 \end{tabular} &\begin{tabular}{c| c} 
 $T \ll \Theta_{D}$ & $T \gg \Theta_{D}$\\ 
 \hline
 $1/\tau_{\text{te}} \sim T^{3}$ &$1/\tau_{\text{te}} \sim T^{0} $ \\ 
 \end{tabular}\\  
 \end{tabular} \\ 
 \hline 
  \end{tabular}
  \end{center}
\label{tab: thermoelectric_scattering}
\end{table}
\end{widetext}

\subsection{Seebeck Coefficient}
As discussed earlier that the Seebeck coefficient is the ratio of the thermoelectric and electrical response functions. Thus, to compute it, we require $\alpha(z,T)$ and $\sigma(z,T)$.\\
The $\sigma(z,T)$, known as electrical conductivity can be computed with the following relation which relates the electrical conductivity with the memory function\cite{gotze_72, navinder_16}.
\bea
\sigma(z,T) = i  \frac{\chi_{0}}{z+M(z,T)},
\label{eqn: electrical}
\eea
where $\chi_{0}$ is the static electric current-electric current correlation function and $M(z,T)$ is the electrical memory function.\\

In the reference (\cite{navinder_16}), it has been derived in detail that the imaginary part of the electrical memory function for the case of the electron-impurity interaction is written as
\bea \nonumber
M''(\omega, T) &=& \frac{N_{\text{imp}} U^2 k_{F}^4}{6 \pi^3 \chi_{0}} \int_{0}^{\infty} d\eta \frac{1}{x} \left( \frac{1}{e^\eta +1} - \frac{1}{e^{\eta + x} +1} \right).\\
\label{eqn: E_MF_impurity}
\eea

In the zero frequency limit, it can be expressed as
\bea \nonumber
M''(T) &=& \frac{N_{\text{imp}} U^2 k_{F}^4}{6 \pi^3 \chi_{0}} \int_{0}^{\infty} d\eta  \frac{e^{\eta}}{(e^{\eta} +1)^2}.\\
\label{eqn: E_MF_dc_impurity}
\eea

Substituting the Eq.(\ref{eqn: E_MF_dc_impurity}) in Eq.(\ref{eqn: electrical}) by taking the zero frequency limit of $\sigma(z, T)$ ($ = \frac{\chi_{0}}{M''(T)}$), we find that the electrical conductivity shows temperature independent behavior in case of the electron-impurity interaction. Also, the thermoelectric conductivity in the former case shows temperature independent behavior (Eq.(\ref{eqn:dc_ETC})) in the zero frequency case. Using these behaviors in Eq.(\ref{eqn: seebeck}), we see that the Seebeck coefficient is temperature independent in case of the electron-impurity interaction.

Similarly in the case of the electron-phonon interaction, the electrical memory function in the finite frequency case and in the zero frequency limit can be written as\cite{gotze_72, bhalla_16, navinder_16}
\bea \nonumber
M''(\omega, T)&=& \frac{N}{24 \pi^3 m_{i} \rho_{F}^2} \left( \frac{q_{D}}{\Theta_{D}}\right)^6 \frac{T^5}{\chi_{0}} \int_{0}^{\Theta_{D}/T} dy \frac{y^4}{(e^{y}-1)} \\ \nonumber
&& \left\lbrace \frac{x-y}{e^{x-y} -1} \frac{e^{x}-1}{x} + (\text{terms with $\omega \rightarrow -\omega$}) \right\rbrace,\\
\label{eqn: E_MF_phonon}
\eea
and
\bea \nonumber
M''(T)&=& \frac{N}{12 \pi^3 m_{i} \rho_{F}^2} \left( \frac{q_{D}}{\Theta_{D}}\right)^6 \\
&& \times \frac{T^5}{\chi_{0}} \begin{cases}
  124.4,& \textnormal{at $T\ll \Theta_{D}$} \\
  \frac{1}{4}\left(\frac{\Theta_{D}}{T} \right)^4,&\textnormal{at $T\gg \Theta_{D}$},
  \end{cases} 
\label{eqn: E_MF_dc_phonon}
\eea
\begin{widetext}
\begin{table}[t]
\caption{The electrical scattering rate due to the electron-impurity and the electron-phonon interactions in different frequency and temperature domains.}
\begin{center}
  \begin{tabular}{|c| c| c| c| c|} 
 \hline
 \multicolumn{3}{|c|}{\textbf{Electron-impurity interaction}}\\
 \hline
 $\omega = 0$ & $\omega \gg T$ & $\omega \ll T$\\ 
  \hline 
  $1/\tau \sim T^{0}$ & \begin{tabular}{c} 
 $1/\tau \sim \frac{T}{\omega} \log 2$ 
 \end{tabular} 
 & $1/\tau \sim T^{0}$ \\
  \hline
   \multicolumn{3}{|c|}{\textbf{Electron-Phonon interaction}}\\
 \hline
 $\omega = 0$ & $\omega \gg \omega_{D}$ & $\omega \ll \omega_{D}$ \\ 
 \hline
 \begin{tabular}{c | c} 
  $T \gg \Theta_{D}$ & $T \ll \Theta_{D}$\\ 
 \hline
 $1/\tau \sim T$ &$1/\tau \sim T^{5} $ \\ 
 \end{tabular} & \begin{tabular}{c| c} 
  $\omega \gg T$ & $\omega \ll T$\\  
 \hline
 \begin{tabular}{c| c} 
 $T \gg \Theta_{D}$ & $T \ll \Theta_{D}$\\ 
 \hline
 $1/\tau \sim T$ &$1/\tau \sim T^{5} $ \\ 
 \end{tabular} &\begin{tabular}{c} 
  $T \gg \Theta_{D}$ \\ 
 \hline
 $1/\tau \sim T$ \\ 
 \end{tabular} \\ 
\end{tabular} & \begin{tabular}{c| c} 
  $\omega \gg T$ & $\omega \ll T$\\ 
 \hline
 \begin{tabular}{c} 
  $T \ll \Theta_{D}$ \\ 
 \hline
 $1/\tau \sim T^{6} \frac{e^{\omega/T}}{\omega}$ \\ 
 \end{tabular} &\begin{tabular}{c| c} 
 $T \ll \Theta_{D}$ & $T \gg \Theta_{D}$\\ 
 \hline
 $1/\tau \sim T^{5}$ &$1/\tau \sim T $ \\ 
 \end{tabular}\\  
 \end{tabular} \\ 
 \hline 
  \end{tabular}
  \end{center}
\label{tab:phonontable}
\end{table}
\end{widetext}
respectively. Substituting this zero frequency electrical memory function in Eq.(\ref{eqn: electrical}), we find that the electrical conductivity in this limit, $\sigma(T) = \frac{\chi_{0}}{M''(T)}$, shows $T^{-5}$ behavior and $T^{-1}$ behavior in the low and the high temperature regimes i.e. $T\ll\Theta_{D}$ and $T\gg \Theta_{D}$ respectively.
On the other hand, we have discussed that the thermoelectric conductivity $\alpha(T)$ shows $T^{-3}$ and a temperature independent behavior in the low and the high temperature regimes respectively. Substituting these into Eq.(\ref{eqn: seebeck}), the Seebeck coefficient in the electron-phonon interaction shows temperature dependence as
follows
\bea
S(T) &\propto& \begin{cases}
  T^{2},& \textnormal{at $T\ll \Theta_{D}$} \\
 T,&\textnormal{at $T\gg \Theta_{D}$}.
  \end{cases}
\eea
In the finite frequency case, using the definition of the Seebeck coefficient (Eq,.(\ref{eqn: seebeck})) $S(z,T)$ is written as
\bea
S(z,T) &=& \frac{1}{T}\frac{\chi_{Q}^{0}(T)}{\chi_{0}} \frac{z+ M(z,T)}{z+M_{Q}(z,T)}.
\eea
Thus, the real part of the Seebeck coefficient becomes 
\bea \nonumber
\text{Re}[S(\omega,T)] &=& \frac{1}{T}\frac{\chi_{Q}^{0}(T)}{\chi_{0}} \frac{\omega^2 + M''(\omega,T)M''_{Q}(\omega,T)}{\omega^2 + (M''_{Q}(\omega,T))^2}.\\
\label{eqn: seebeck_realpart}
\eea
Substituting the imaginary part of the electrical and the thermoelectric memory function (Eqs.(\ref{eqn:memoryimpurity_final}) and (\ref{eqn: E_MF_impurity}) for impurity case and Eqs.(\ref{eqn: TE_MF_phonon}) and (\ref{eqn: E_MF_phonon}) for phonon case) in Eq.(\ref{eqn: seebeck_realpart}), we can discuss the frequency variation of the Seebeck coefficient for the case of the electron-impurity and electron-phonon interactions in the next section. In the case of metal, the dynamic behavior of various transport coefficients is mainly governed by the imaginary part of the memory function or the scattering rate\cite{gotze_72}. Owing to this fact, we ignore the contribution of the real part of the memory function in this discussion. \\

\section{Results}
\label{sec: results}
In this section, we have presented the results for the imaginary part of the thermoelectric memory function and the corresponding thermoelectric coefficient in different temperature and frequency domains.
\figimpurityE

In Fig. \ref{fig: ac_impurity}, we plot the imaginary part of the normalized thermoelectric memory function (Eq.(\ref{eqn:memoryimpurity_final})) $M''_{Q}(\omega,T)/M''_{0}$, where $M''_{0} = \frac{2 k_{F}^4 m N_{\text{imp}} U^2 }{3 \pi^4 N_{e} }$, due to the electron-impurity interaction as a function of $\omega$ and at different temperatures. Here we observe that the thermoelectric memory function at low frequency i.e. $\omega \ll T$ shows frequency and temperature independent behavior (as shown in fig.(\ref{fig: ac_impurity_full})). In the intermediate regime, it decays with the increase of the frequency (as shown in Fig.\ref{fig: ac_impurity_half}). Also it increases with the increase of the temperature. Finally, at high frequencies i.e. $\omega \gg T$, it saturates to constant value.
\figimpurityF

In Fig. \ref{fig: seebeck_ac_impurity}, the real part of the normalized Seebeck coefficient $\text{Re}[S(\omega,T)]/S_{0}$ for the case of electron-impurity interaction is shown as a function of $\omega/\omega_{0}$, ($\omega_{0} = \frac{2 k_{F}^4 m N_{\text{imp}} U^2 }{3 \pi^4 N_{e} }$) and at different temperatures. Here, we observe that in the low frequency regime, it is constant and then starts decrease with the rise of the frequency and shows a dip at a certain frequency (Fig.\ref{fig: ac_seebeck_impurity_full}). Then, in the high frequency regime, it saturates again to the constant value. Also, with the rise of temperature, $\text{Re}[S(\omega,T)]/S_{0}$ increases in the low frequency regime and becomes independent of the temperature in the high frequency regime. This behavior can be understood from Eq.(\ref{eqn: seebeck_realpart}) as follows:

In the high frequency regime, Eq.(\ref{eqn: seebeck_realpart}) can be written as
\bea \nonumber
\text{Re}[S(\omega,T)] &\approx& \frac{1}{T}\frac{\chi_{Q}^{0}(T)}{\chi_{0}} \\
&\approx& \textnormal{constant}.
\eea
This feature is dipicted in Fig.(\ref{fig: ac_seebeck_impurity_full}).

Now, in the low frequency regime i.e. $\omega \rightarrow 0$, Eq.(\ref{eqn: seebeck_realpart}) is approximated as
\bea
\text{Re}[S(\omega,T)] &\approx& \frac{1}{T}\frac{\chi_{Q}^{0}(T)}{\chi_{0}} \frac{M''(\omega,T)}{M''_{Q}(\omega,T)}.
\eea
Within this low frequency regime, the thermoelectric and the electrical memory function\cite{navinder_16} shows saturation behavior in frequency at a finite temperature, hence the Seebeck coefficient saturates (as shown in Fig.\ref{fig: ac_seebeck_impurity_half} below the value $0.01$.).

In Fig.(\ref{fig: ac_phonon_freq}), we plot the frequency and the temperature dependent normalized imaginary part of the thermoelectric memory function $M''_{Q}(\omega, T)/M''_{0}$ for case of the electron-phonon interaction, where $M''_{0} = \frac{N m q_{D}^6}{6\pi^5 m_{i} N_{e} \rho_{F}^2 \Theta_{D}}$. Here, we keep the Debye temperature $\Theta_{D}$ fixed at $300$K i.e. $0.026$eV and look at the frequency dependence at different temperatures. We observe that the thermoelectric memory function shows frequency variation in the region from $0.02$eV to $0.2$eV. While in other regions i.e. at extremely low $\omega \ll 0.02$eV and high frequency $\omega \gg 0.2$eV regimes, it shows frequency independent behavior (Fig. \ref{fig: ac_phonon_memory_full} and \ref{fig: ac_seebeck_impurity_half}). Along with the frequency character, we also observe the temperature behavior. In throughout the frequency region, it increases with the increase of the temperature (Fig. \ref{fig: ac_phonon_memory_full} and \ref{fig: ac_seebeck_impurity_half}).

\figphononA
Now, in Fig. \ref{fig: dc_phonon} we plot the imaginary part of the thermoelectric memory function in the zero frequency limit as a function of temperature. Here, we consider different values of the Debye temperature such as $200$, $300$ and $400$K. It is found that $M''_{Q}(T)/M''_{0}$ first increases with the increase of temperature and then saturates to a constant value at temperature above the Debye temperature.

\figphononB
\figphononC
In Fig. \ref{fig: seebeck_ac_phonon}, we plot the real part of the frequency and temperature dependent normalized Seebeck coefficient $\text{Re}[S(\omega,T)]/S_{0}$ with $\omega/\omega_{0}$ (using Eq.(\ref{eqn: seebeck_realpart})) at different temperatures. Here $\omega_{0}$ $\left(= \frac{N m q_{D}^6}{6\pi^5 m_{i} N_{e} \rho_{F}^2 \Theta_{D}}\right)$ is the scaling parameter. We have kept the Debye temperature fixed at $300$K. In Fig. \ref{fig: seebeck_ac_phonon_full}, we observe that $\text{Re}[S(\omega,T)]/S_{0}$ is independent of the frequency and the temperature at the high frequency regime (i.e. $\omega  \gg \omega_{D}$ as shown in the regime right to the dashed line within a plot). In contrast, there is strong frequency and temperature dependence at the low frequency regime. To elaborate the low frequency regime, we replot the real part of the Seebeck coefficient in Fig. \ref{fig: seebeck_ac_phonon_half}. Here we find that the later increases with the increase in frequency. While with the rise in temperature, the magnitude of $\text{Re}[S(\omega,T)]/S_{0}$ reduces. The saturation at high frequencies can be understood from formula (Eq.(\ref{eqn: seebeck_realpart})) as explained above. Also the suppression of the normalized Seebeck coefficient with the increase in temperature can be understood by recognizing the enhanced scattering of quasiparticles at higher temperature. At very low frequency, we show the same plot at temperature $300$K within the inset of Fig. \ref{fig: seebeck_ac_phonon_half}. We can see from the inset that near the zero frequency, $\text{Re}[S(\omega,T)]/S_{0}$ approaches to the constant value.\\
\figphononD

For the zero frequency case, $\text{Re}[S(T)]/S_{0}$ (using Eq.(\ref{eqn: seebeck_realpart}))is plotted as a function of temperature at different Debye temperatures such as $200$, $300$ and $400$K in fig.(\ref{fig: seebeck_dc_phonon}). It is observed that $\text{Re}[S(T)]/S_{0}$ increases linearly with the rise of temperature. Also, its linear behavior is more pronounced at the temperature more than the Debye temperature. This linear behavior feature is in accord with the result calculated by Boltzmann approach and with the experimental findings. However, at very low temperature ($T \ll \Theta_{D}$), it is quadratic in temperature. Experimentally, this regime is dominated by the phonon drag effects\cite{ziman_book} and these are not considered in the present formalism.

\section{Conclusion}
\label{sec: conclusion}
Making highly efficient thermoelectric devices, one needs materials with large figure of merit ($ZT = S^2\sigma T / \kappa$). As discussed earlier, one possible root to increase $ZT$ is to look beyond the static limit and look for the frequency dependent case. In this connection the understanding of the frequency dependence of the Seebeck coefficient $S(\w, T)$ is extremely important and is attempted here.

As discussed in many of the previous works, the memory function approach has many advantages over the much celebrated Bloch-Boltzmann approach in studying the dynamical behavior of various transport properties. Here we for the first time attempted a theoretical studies of thermoelectric phenomenon within the memory function formalism.

In this work, we find that the Seebeck coefficient in the zero frequency limit shows the temperature independent behavior in the case of the impurity interaction, whereas in the case of electron-phonon interaction, it shows linear temperature behavior in the high temperature regime (i.e. $T\gg\Theta_{D}$) which is in accord with the Boltzmann results (the famous Mott formula) and experimental results\cite{kamran_book, ziman_book}. In the low temperature regime (i.e. $T \ll \Theta_{D}$), the Seebeck coefficient shows the quadratic temperature dependence.

Looking at the frequency dependence of $S(\omega,T)$, we found that in the electron-impurity scattering case, the $S(\omega,T)$ decays with the increase in frequency in the low frequency regime and then after passing through a minimum, it becomes constant in the high frequency regime. Contrary to that in the case of the electron-phonon interaction, it rises with the frequency in the regime where $\omega \ll \omega_{D}$ and in the opposite case i.e. $\omega \gg \omega_{D}$, it saturates. These new predictions insure that the phonon interaction plays an important role in the dynamical behavior of the Seebeck coefficient and hence can help in improving the figure of merit. More precisely, a thermoelectric material with stronger electron-phonon interaction and operating at a certain finite frequency should have a larger contribution to its figure of merit from its Seebeck coefficient. Although, this new route is important to study the properties of the thermoelectric materials, but still lot of questions such as the contributions of the Umklapp processes and phonon drag effects, etc. has to be addressed. It is known that at low temperature phonon drag effects are important but how these effects modify the finite frequency Seebeck coefficient remains an open problem.

\appendix
\section{Static Thermoelectric Correlation Function}
\label{app: static_derivation}
The static thermal current-electric current correlation is defined as
\be
\chi_{Q}^{0}(T) = \frac{1}{3 T} \sum_{\textbf{k}} (\epsilon_{\textbf{k}} - \mu) v_{\textbf{k}}^2 f_{\textbf{k}} (1-f_{\textbf{k}}).
\label{eqn: app1}
\ee
Converting the summation over electron momentum into energy integral and substituting $\frac{\epsilon_{\textbf{k}} - \mu}{T} = \eta$, Eq.(\ref{eqn: app1}) becomes
\bea \nonumber
\chi_{Q}^{0}(T) &=& \frac{T k_{F}^3}{3 m} \frac{1}{2\pi^2} \int_{0}^{\infty} d\eta \frac{\eta e^{\eta}}{(e^{\eta}+1)^2}. \\
&=& T \frac{N_{e}}{3 m} \log2.
\label{eqn: static_expression}
\eea
This shows that the static thermal current-electric current correlation varies linearly in temperature.

\end{document}